\documentclass[journal,letterpaper,twocolumn,final,10pt]{IEEEtran}
\usepackage[notrig]{physics}
\usepackage{amsmath} 
\usepackage{ulem} 
\usepackage{mathtools}
\usepackage{hyperref}
\usepackage{amssymb}
\usepackage{comment}
\usepackage{algorithm}
\usepackage{algorithmic}
\usepackage{color}

\newcommand{\testest}{\mbox{$
		\begin{array}{c}
			\stackrel{ \stackrel{\textstyle H_{\hat{k}}}{\textstyle >} }{
				\stackrel{\textstyle <}{\textstyle H_0} }
		\end{array}
		$}}
        \newcommand{\test}{\mbox{$
		\begin{array}{c}
			\stackrel{ \stackrel{\textstyle H_{{K\geq k}}}{\textstyle >} }{
				\stackrel{\textstyle <}{\textstyle H_{k-1}} }
		\end{array}
		$}}
        
\begin{document}
	
	\title{An Information-Theoretic Detector for\\ Multiple Scatterers in SAR Tomography}
	
	\author{Pia Addabbo, \IEEEmembership{Senior Member, IEEE},
		Diego Reale, \IEEEmembership{Member, IEEE},
        Antonio Pauciullo, \\
		Gianfranco Fornaro, \IEEEmembership{Fellow Member, IEEE},
		 and Danilo Orlando,~\IEEEmembership{Senior Member, IEEE}
		 \thanks{This work has been submitted to the IEEE for possible publication. Copyright may 
		 be transferred without notice, after which this version may no longer be accessible}
		\thanks{Pia Addabbo is with Universit\`a degli Studi ``Giustino Fortunato", viale Raffale Delcogliano, 12, 82100 Benevento, Italy. Email: {\tt p.addabbo@unifortunato.eu}}
		\thanks{Diego Reale, Antonio Pauciullo and Gianfranco Fornaro are with National Research Council - Institute for the Electromagnetic Sensing of the Environment (CNR-IREA), via Diocleziano 328, 80124, Naples, Italy.
 E-mail: \{\tt{reale.d, pauciullo.a, fornaro.g\}@irea.cnr.it}.}
		\thanks{Danilo Orlando is with the Department of Information Engineering, University of Pisa, via Girolamo Caruso 16, 56122, Pisa, Italy.
  Email: {\tt danilo.orlando@unipi.it}}}

\maketitle

\begin{abstract}
Persistent scatterer interferometry and Synthetic Aperture Radar (SAR) Tomography are powerful tools 
for the detection and time monitoring of persistent scatterers.
They have been proven to be effective in urban scenarios,
especially for buildings and infrastructures 3-D reconstruction and monitoring of deformation.  
In urban areas, occurrence of layover leads to the presence of multiple contributions within the same image pixel from scatterers located at different heights. In the context of SAR Tomography, this problem can be addressed by considering a multiple hypothesis test to detect the presence of feasible multiple scatterers \cite{Pauciullo2012, budillon2015glrt}. In the present paper, we consider this problem in the framework of the information theory and exploit the theoretical tool, developed in \cite{addabbo2021adaptive}, to design a one-stage adaptive architecture for multiple hypothesis testing
		problems in the context of SAR Tomography. Moreover, we resort to the compressive sensing approach for the estimation of the unknown parameters under each hypothesis. This architecture has been verified on both simulated as well as real data also in comparison with suitable counterparts.
	\end{abstract}
	

\IEEEpeerreviewmaketitle

\section{Introduction}
\IEEEPARstart{S}{ynthetic} Aperture Radar (SAR) Interferometry (InSAR) and,
in particular, Differential InSAR (DInSAR) have given an
impetus to the use of SAR for monitoring environmental
hazards \cite{ferretti2001permanent, berardino2002new, ferretti2000nonlinear}. Modern DInSAR methods, known as Advanced
DInSAR (A-DInSAR) techniques, exploit stacks of
SAR images acquired on repeated orbits, with time intervals
of months or years, to accurately monitor slow ground
displacements. Persistent Scatterer Interferometry (PSI) belongs to the class of A-DInSAR techniques. PSI techniques rely on the fact that scatterers with a sufficiently high temporal
coherence, i.e., Persistent Scatterers (PS), can be identified
at the highest resolution. They are typically located above
man-made structures, e.g., buildings or infrastructures, or on
exposed rocks. The PSI shows the capabilities of multitemporal/multibaseline interferometry analysis. In fact, the temporal 
coherence indicator can be exploited as a detector of PSs and allows estimating their position and deformation. Within the framework of 
multibaseline interferometric stacking, SAR Tomography (TomoSAR) 
provides a different perspective and processing strategy with respect 
to PSI. In fact TomoSAR technique, validated both on airborne 
\cite{reigber2000first} as well as spaceborne SAR data 
\cite{FornSold2003, Fornaro2005}, allows us to frame SAR 
interferometry into 3D or more multidimensional SAR imaging context. 
Under this operating approach, PSs appear as scatterers focused along the cross 
slant range direction. This focusing technique can be extended to the 
4D case, which scans the cross slant range-velocity space \cite{4Dfornaro, Zhu2010}, or even 
higher dimensions \cite{reale2013extension, Zhu2011warp}.
 
In  this context, the goal is to derive a simple and effective
detection scheme that can improve the detection capability
of man-made targets in areas with low Signal-to-Noise Ratio (SNR), 
such as rural
areas, at the maximum resolution. The problem of jointly detecting
multiple point-like targets is a difficult task since target 
parameters and, more importantly, the target number
are unknown and have to be estimated from the available data. 
Therefore,
unlike conventional detection problems that include two
hypotheses, i.e., the noise-only (or null) hypothesis and the
signal-plus-noise (or alternative) hypothesis, this lack of a
priori information naturally leads to multiple alternative hypotheses 
(the null hypothesis remains the conventional one).
Several strategies have been proposed to address this problem over 
the years. A detection scheme grounded on Generalized 
Likelihood Ratio Test (GLRT) for the detection of single scatterers in SAR Tomography was firstly introduced in \cite{DeMaio2009}. Its extension to the multiple hypothesis for the detection of multiple PSs in a 
tomographic cell test was subsequently proposed in  
\cite{Pauciullo2012}. This detection scheme, based on the cancellation of the first stronger scatterer and therefore referred to as Sequential GLRT with Cancellation (SGLRT-C), implements of a sequence of two decision tests checking for the presence of at maximum one scatterer, thus also limiting the computational requirements. 
However, the cancellation of the first scatterers limits the super-resolution capability of the detection scheme, i.e. the possibility to detect scatterers below the Rayleigh resolution.

To overcome super-resolution limitations of the SGLRT-C in \cite{Pauciullo2012}, a GLRT detector, based on support estimation 
(Sup-GLRT), for multiple scatterers detection in TomoSAR, has been proposed in 
\cite{budillon2015glrt}. The Sup-GLRT also combines a sequence of two detection stages, with the latter scanning the entire 2-dimensional subspace spanned by the possible two scatterers, thus allowing also achieving super-resolution capabilities. Moreover, in principle, the test could be extended to an arbitrary ($>2$) number of interfering scatterers, although this is practically limited by increasing computational efforts. Recently, a solution aimed ar reducing the computational cost has been proposed in \cite{fast_supglrt} where a fast Sup-GLRT implementation reducing the computational cost from combinatorial, with respect to all the possible positions in the search grid to linear, at the expense of a reasonable small performance loss, has been preserved.
At the design stage, Sup-GLRT exploits a sparsity 
assumption for the signal model along the elevation (cross slant range) 
direction, which is often verified in urban areas. Compressed Sensing 
(CS) is a powerful technique based on the sparsity nature of the 
signal under investigation \cite{candes2006robust} and that can be 
applied to TomoSAR systems \cite{zhu2010tomographic}, 
\cite{budillon2009sar}. CS reduces the required number of 
measurements and enhances the elevation resolution
\cite{budillon2010three}, \cite{zhu2011super}. 
{\color{black} A recent and efficient approach \cite{rs17142334} proposes a fast GLRT-based  method that bypasses the grid-based search by performing a continuous
optimization of an objective function to find suitable estimates
of the unknown parameters. To this end, the method uses a quasi-Newton descent algorithm that is initialized and executed under each hypothesis. The experimental results show a superior (detection and estimation) performance with respect to the fast GLRT-based algorithm
proposed in \cite{fast_supglrt}.}
{\color{black} Another interesting approach allowing for super-resolution in 
TomoSAR is described in \cite{9281339}, where model order selection rules
\cite{Stoica1} and suitable regularization parameters are incorporated into
spectral (or direction) estimation algorithms such as the multiple signal classification
and the estimation of signal parameters via rotational invariant techniques.
In addition, the authors investigate different methodologies for the proper
estimation/selection of the related parameters.}

{\color{black}
Most of the mentioned solutions are based on a grid of points used
to search the most likely values for the unknown parameters. However, 
as also explained in Section IV, in case of mismatch a performance degradation
can occur. To overcome this drawback, gridless solutions were introduced such as
in \cite[and references therein]{10734323,rs17142334}. Such solutions are grounded
on gradient descent algorithms, atomic norm minimization, particle
swarm optimization, and so on.
}
However, both the SGLRT-C and Sup-GLRT  
experience some issues related to the thresholds setting. Specifically, due to the presence of many alternative hypotheses, a set of thresholds must be considered to select the hypotheses. In fact, existent solutions 
are based on multiple stages systems which make difficult and time 
demanding the selection of a suitable threshold to ensure a 
preassigned Probability of False Alarm (PFA) as well as a
reliable probability of correct classification.

In this paper, we exploit the
elegant and systematic framework developed in \cite{addabbo2021adaptive} that is based on the Kullback-Leibler
Information Criterion (KLIC) \cite{kullback1951information}. 
This design framework considers
multiple hypotheses testing problems formed by many alternative
hypotheses and only one null hypothesis. Thus, it allows us to
suitably handle the lack of information related to the number
of PSs. In this context, we design a decision scheme capable of detecting
an unknown number of PSs, and estimating their geophysical parameters of interest, i.e. positions and possible displacement rate, in the following referred to as velocity, by using an unique threshold for all the 
hypotheses. Moreover, we incorporate the CS paradigm within
this framework by resorting to a sparse formulation of the related estimation problems. This choice is dictated by the fact that
the conventional maximum likelihood approach used in
\cite{addabbo2021adaptive} would lead to prohibitive computational
requirements.
Thus, in the context of the TomoSAR, we
merge the CS paradigm and information-theoretic design criterion
to come up with a detection architecture, referred to as KLIC-based detector (KLIC-D) that uses a unique
threshold to deal with multiple hypotheses unlike existing approaches.
It is worth pointing out that the threshold setting for the proposed
architecture does not become involved as the number of hypotheses
increases. In fact, compared to the Sup-GLRT, the proposed approach offers a significant advantage: the use of a single adaptive threshold (along with a possible design parameter as explained in the next sections) valid for all alternative hypotheses greatly simplifies the design stage while ensuring a constant false alarm rate (CFAR), even when the number of scatterers is unknown. This architecture
represents the main technical novelty of this contribution and
appears here for the first time (at least to the best of the authors’
knowledge).
The performance analysis is conducted on both synthetic and real data and a comparison with Sup-GLRT is also carried out. The numerical examples highlight that the proposed architecture returns similar performances to that of the Sup-GLRT without using involved multi-stage decision statistics. As consequence, from a practical point of view, the proposed architecture is more manageable than the considered counterpart especially in the presence of a large number of alternative hypotheses.

The remainder of this paper is organized as follows. In the next section, we provide a formal definition of the signal model and detection problem at hand. In Section \ref{sec_SLdetectors}, we derive the adaptive KLIC-D decision scheme and describe the estimation procedures for the unknown parameters. In Section \ref{sec_performance_sim}, we assess the nominal behavior of the proposed detection algorithm by using synthetic data, whereas in Section \ref{sec_performance_real}, we present the detection results over real COSMO-SkyMed data. Finally, in Section \ref{sec_conclusions}, we draw concluding remarks and trace the route for future research lines.

\section*{Notation}
In the sequel, vectors and matrices 
are denoted by boldface lower-case and upper-case letters, respectively.
Symbols $(\cdot)^T$, and $(\cdot)^\dag$ denote transpose and conjugate transpose (Hermitian) operators, respectively.
Symbol $\|\cdot\|$ denotes the Euclidean norm of a vector. 
$\mathbb{R}$ is the set of real numbers, and $\mathbb{R}^{N\times M}$ is the Euclidean space of $(N\times M)$-dimensional 
real matrices (or vectors if $M=1$).
$\mathbb{C}$ is the set of complex numbers, and $\mathbb{C}^{N\times M}$ is the Euclidean space of $(N\times M)$-dimensional 
complex matrices (or vectors if $M=1$). If
$\mathbf{a}$ is an $N$-dimensional vector, then \text{diag}$(\mathbf{a})$
is an $N\times N$ diagonal matrix whose nonzero entries are the elements
of $\mathbf{a}$.
Symbol $\mathbf{I}$ stands for the identity matrix of proper size. 
Given a generic hypothesis denoted by $H$, $\bar{H}$ is the complement of $H$.
The acronym PDF stands for Probability Density Function and the conditional PDF of a random variable $x$ given another random variable $y$ is denoted by $f(x|y)$. 
Finally, we write $\mathbf{x}\sim\cal{C}\cal{N}_{\textit{N}}(\boldsymbol{\mu}, \mathbf{M})$ if $\mathbf{x}$ is a  complex circular $N$-dimensional normal vector with mean $\boldsymbol{\mu}$ and positive definite covariance matrix $\mathbf{M}$.

	
\section{Signal model and Problem Formulation}
\label{sec_sig_model}

\begin{figure}[htp!]
    \begin{center}
        \includegraphics[scale=0.35]{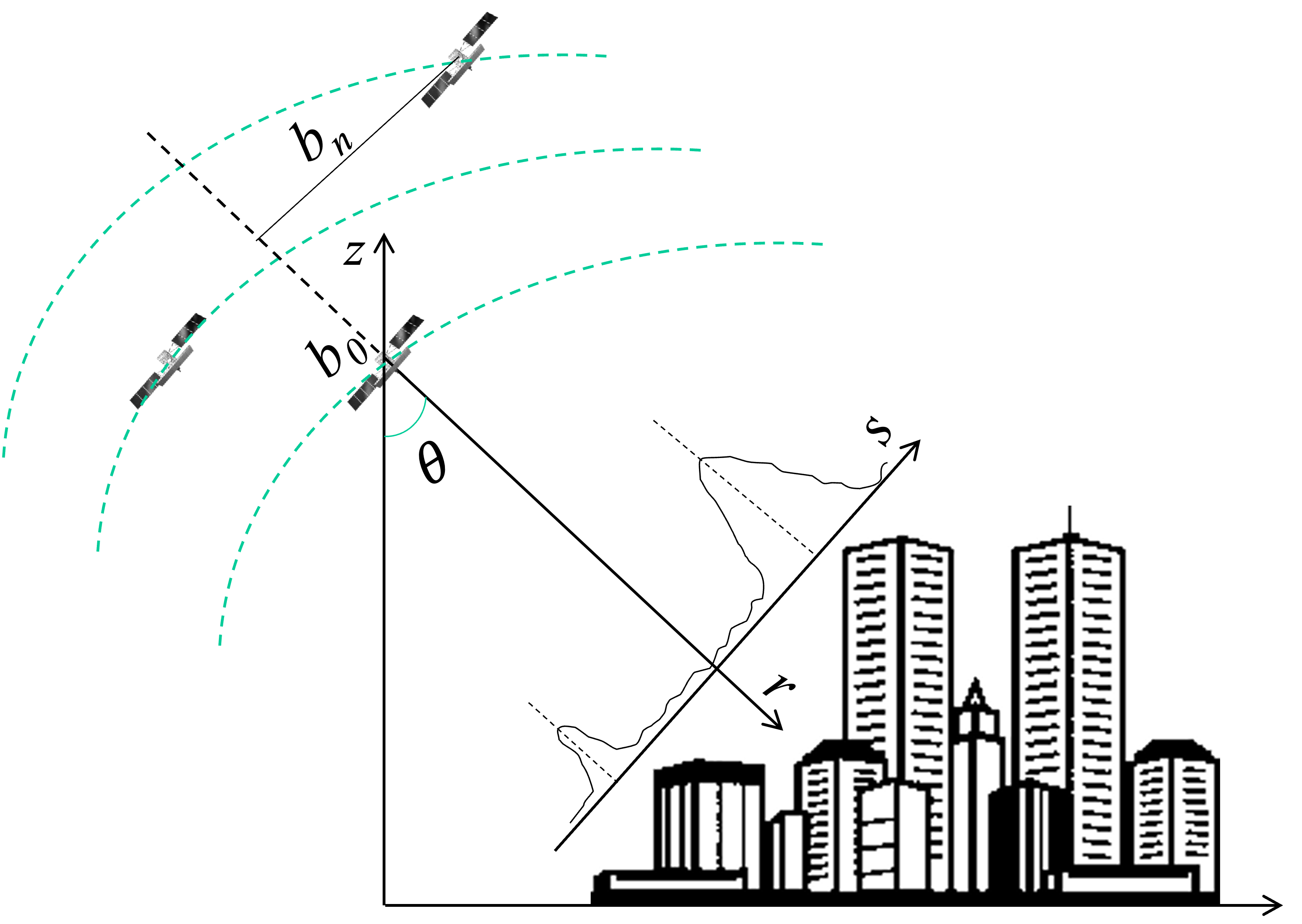}
        \caption{Multi-temporal tomographic SAR geometry, with the radar acquiring from different positions separated by $b_n$, $n=1,...,N$; $\theta$ is the look angle, $s$ is the elevation direction (orthogonal to the line-of-sight $r$).}
        \label{fig:figure00}
    \end{center}
\end{figure}

Let us consider $N$ SAR images acquired at different time epochs $t_n, n=1,\dots,N$. 
The received signal in any image pixel, after coregistration with respect to a given (reference) image and atmospheric effects compensation, can be modeled as the superposition of elementary contributions distributed along the elevation direction $s$, also referred to as slant-height, orthogonal to radar line-of-sight $r$, and related to the height $z=s\sin(\theta)$, with $\theta$ being the incidence angle, see Figure \ref{fig:figure00}, \cite{4Dfornaro}:
\begin{equation}
    x_n = \int_{\Delta s} \gamma(s)e^{-j\frac{4\pi}{\lambda}d(s,t_n)} e^{-j2\pi\xi_n s}ds,
    \label{eq_xn}
\end{equation}
where:
\begin{itemize}
    \item $\gamma(s) \in \mathbb{C}$ is the distribution of the backscattering coefficient along $s$. We assume $\gamma(s)$ to be constant over the time;
    \item $d(s,t_n) \in \mathbb{R}$ is the distance variation at the different acquisition epochs;
    \item $\xi_n = \frac{2b_n}{\lambda r_0}$ represents the spatial frequency with $b_n$ being perpendicular baseline of the $n$-th acquisition, $r_0$ the distance of the target from the reference track, and $\lambda$ the system wavelength;
    \item $\Delta s$ is the extent of the scene in elevation direction.
\end{itemize}

The exponential deformation term in \eqref{eq_xn} can be expanded as the 2D-Fourier transform of a suitable function $g(s,v,l)$ \cite{reale2013extension}, i.e., 
\begin{equation}
    e^{-j\frac{4\pi}{\lambda}d(s,t_n)} = \int_{\Delta v}\int_{\Delta l} g(s,v,l) e^{-j2\pi(\eta_nv+\zeta_nl)}dvdl,
    \label{eq_armonics}
\end{equation}
where $v$ and $l$, associated with the Fourier mate variables $\eta_n = \frac{2t_n}{\lambda}$ and $\zeta_n = \frac{2T_n}{\lambda}$, with $T_n$ being the temperatures at the acquisition epochs, 
play the role of velocity and of thermal dilation coefficients. Substituting \eqref{eq_armonics} in \eqref{eq_xn}, we have 
\begin{equation}
    x_n = \int_{\Delta s}\int_{\Delta v}\int_{\Delta l} \gamma(s,v,l)e^{-j2\pi(\xi_ns+\eta_nv+\zeta_nl)}ds dv dl,
    \label{eq_xnF}
\end{equation}
with $\gamma(s,v,l)=\gamma(s)g(s,v,l)$.

Let us rearrange the received signal in vector form and consider $\mathbf{x}\in\mathbb{C}^{N\times 1}$, the vector collecting, for each
pixel, the measured complex data from the $N$ images. Assuming concentrated scattering in \eqref{eq_xnF} and considering the noise component, the signal vector can be expressed as the superposition of an unknown number, say $K\le K_{\max}$, of fully coherent scatterers \cite{8326727}:
\begin{equation}
    \mathbf{x}=\sum\limits_{k=1}^K g_k \mathbf{a}(\mathbf{p}_k) + \mathbf{w},
    \label{eq_pixelmodel}
\end{equation}
where:
\begin{itemize}
    \item $\mathbf{w} \sim \cal{C} \cal{N}_{\textit{N}} (\mathbf{0},$ $\sigma^{2} \mathbf{I})$ is the disturbance vector with $\sigma^{2} > 0$;
    \item ${g}_k\in\mathbb{C}, k=1,\dots,K,$ is the unknown $k$th complex backscattering coefficient;
    \item $\mathbf{p}_k=\left[s_k,v_k,l_k\right]^T\in \mathbb{R}^{3\times 1}, k=1,\dots,K,$ represents the unknown $k$th scatterer position in the parameter space;
    \item $\mathbf{a}({\mathbf{p}}_k)\in\mathbb{C}^{N \times 1}, k=1,\dots,K,$ is the normalized steering vector (i.e., $\|\mathbf{a}({\mathbf{p}}_k)\|=1$). The $n$th component of the $k$th steering vector is given by
   $
   \{\mathbf{a(p}_k)\}_n = \frac{1}{\sqrt{N}}\exp(-j2\pi\boldsymbol{\zeta}^T_n\mathbf{p}_k),
   $
   with $\boldsymbol{\zeta}_n=\left[\xi_n,\eta_n, \zeta_n\right]^T$ being the vector collecting the frequencies associated with the parameters in $\mathbf{p}_k$.
\end{itemize}

Defining the steering matrix $\mathbf{A}_{K}=[ \mathbf{a}(\mathbf{p}_1), ... ,  \mathbf{a}(\mathbf{p}_{K})] \in\mathbb{C}^{N \times K}$, we can rewrite the observation vector according to the following linear model
\begin{equation}
    \mathbf{x}= \mathbf{A}_K\mathbf{g}_K + \mathbf{w},
    \label{alternative_equation_model_}
\end{equation}
where $\mathbf{g}_K=\left[g_1,\dots,g_K\right]^T$ is the backscattering coefficients' vector.

Now, the problem of detecting, pixel by pixel, an unknown number, say $k\in\{1,\ldots, K_{\max}\}$, of scatterers, whose related parameters, say $\mathbf{p}_k$, are unknown, can be formulated as the following multiple hypothesis test
\begin{equation}
    \begin{cases}
        \label{eq_multipleHyp}
        H_0&:  \mathbf{x}=\mathbf{w},\\
        H_1&:  \mathbf{x}= \mathbf{A}_1 {g}_1  + \mathbf{w},\\
        & \hspace{0.7 cm} \vdots \\
        H_{k}&:  \mathbf{x}= \mathbf{A}_{k} \mathbf{g}_{k} + \mathbf{w},\\
        & \hspace{0.7 cm} \vdots \\
        H_{K_{\max}}&:  \mathbf{x}= \mathbf{A}_{K_{\max}} \mathbf{g}_{K_{\max}} + \mathbf{w}.
    \end{cases}
\end{equation}
Finally, for the subsequent developments, it is worth providing the PDF of $\mathbf{x}$, under $H_0$, that is
\begin{equation}
    f_0(\mathbf{x};\sigma^2)=\frac{1}{\pi^N (\sigma^2)^N} \exp\biggl\{ -\frac{\|\mathbf{x}\|^2}{\sigma^2}\biggr\},
    \label{eq_pdf_H0}
\end{equation}
and that under the generic $H_k$ hypothesis, $k=1,\dots,K_{\max}$, namely
\begin{equation}
    f_{1,k}(\mathbf{x};\sigma^2,\mathbf{A}_k,\mathbf{g}_k)=\frac{1}{\pi^N (\sigma^2)^N} \exp\biggl\{ -\frac{\|\mathbf{x}-\mathbf{A}_k\mathbf{g}_k\|^2}{\sigma^2}\biggr\}.
    \label{eq_pdf_H1}
\end{equation}

\section{Multiple scatterers Detection}
\label{sec_SLdetectors}

In this section, we devise a detection architecture for  problem \eqref{eq_multipleHyp}  that is based upon the Penalized Log-Likelihood Ratio Test (PLLRT) \cite{addabbo2021adaptive,9456070}.
Actually, we heuristically modify the PLLRT by replacing the maximum likelihood estimates of the unknown parameters under the multiple alternative hypotheses (e.g., $\mathbf{p}_{1}, ... , \mathbf{p}_{k}$, and $\mathbf{g}_k$ in addition to $k$) with suitable estimates obtained by means of the iterative procedure described below. This choice is dictated by the fact that the maximum likelihood approach leads to intractable mathematics and/or time demanding optimization approaches.
Therefore, we start from the general form for the PLLRT given by
\begin{align}
    \label{eqn:penLRT}
    \max_{k\in\{1,2,\dots,K_{\max}\}} \Biggl\{ &\max\limits_{\sigma^2,\mathbf{A}_k,\mathbf{g}_k}\log[f_{1,k}(\mathbf{x};\sigma^2,\mathbf{A}_k,\mathbf{g}_k )] \Biggr.\\ \nonumber
    &\biggl. - {\max\limits_{\sigma^2}\log[f_0(\mathbf{x};\sigma^2)]} -h(k) \Biggr\}\testest \eta,
\end{align}
where $\hat{k}$ maximizes the left-hand side of the above equation,
$\eta$ is the detection threshold set according to a given PFA, defined as $P_{fa}=P(\text{accept } {H}_{\hat{k}}|{H}_0)$, 
$h(k)$ is a penalty term depending
on the number of unknown parameters of interest \cite{1336828} and, inspired by the generalized information criterion, under the $k$-th alternative hypothesis, it is given by
\begin{equation}
    h(k)= \frac{6k}{2} (1+\rho)
    \label{eq_penalty}
\end{equation}
with $\rho>1$ a constant design parameter.
{\color{black} This parameter is inherited from the generalized information criterion \cite{addabbo2021adaptive,Stoica1}. It represents a degree of freedom that allows for a control of the probability of overestimating the number of scatterers. Actually,
it depends on the performance metrics, the number of samples, 
and the data-generating mechanism itself; more importantly, there does not exist a clear indication of how $\rho$ can be selected \cite[and references therein]{Stoica1}. For this reason, in what follows, this parameter is set by heuristically evaluating the probability of misclassification (see Section \ref{sec_performance_sim} for further details).}	

Solving the joint maximization problem with respect to the unknown 
parameters is involved from a mathematical point of view (at 
least to the best of the authors' knowledge). 
For this reason, we exploit the following two-step procedure. First, we
assume that $\sigma^2$ is known and estimate $\bf A_k$, $k=1,
\ldots,K_{\max}$, by means of a compressive sensing strategy
as described in Subsection \ref{sec_compr}.
Then, we use these estimates to compute the logarithm of the GLRT under 
each hypothesis. Recall that the overall (suboptimum) GLRT is derived under the assumption
that $\sigma^2$ and $\bf g_k$, $k=1,\ldots,K_{\max}$, are unknown.

In fact, once the estimates of the $\mathbf{A}_k$s, say 
$\widehat{\mathbf{A}}_k$, are available, we consider the following optimization problems.
Under $H_1$, we obtain
\begin{align}
	&\max_{\mathbf{g}_k\atop k=1,\ldots,K_{\max}}
    \max_{\sigma^2}\biggl[-N\log(\pi)-N\log(\sigma^2)-\frac{1}{\sigma^2}\|\mathbf{x}-\mathbf{\widehat{{A}}}_{k}\mathbf{{g}}_{k}\|^2\biggr]
    \nonumber
    \\
    &=\max_{\mathbf{g}_k}\biggl[C - N\log(\|\mathbf{x}-\mathbf{\widehat{{A}}}_{k}\mathbf{{g}}_{k}\|^2) \biggr]
    \nonumber
    \\
    &=C-N\log(\bf x^\dag P^\perp_{\widehat{A}_k}x)
	\label{eq_last_comprH1},
\end{align}
where $C=-N\log(\pi)-N+N\log N$ and 
$\mathbf{P}^\perp_{\widehat{\mathbf{A}}
_{k}}=\mathbf{I}-\widehat{\mathbf{A}}_{k}(\widehat{\mathbf{A}}^{\dag}
_{k}\widehat{\mathbf{A}}_{k})^{-1}\widehat{\mathbf{A}}_{k}^{\dag}$ is 
the projection matrix onto the orthogonal complement of the subspace
spanned by the columns of $\mathbf{\widehat{A}}_k$. 
\textcolor{black}{Notice that the maximum likelihood estimate of $\sigma^2$ assuming that 
$\mathbf{A}_k=\widehat{\mathbf{A}}_k$ is given by
\begin{equation}
\widehat{\sigma}_k^2=\frac{1}{N}\| \mathbf{x}
-\widehat{\mathbf{A}}_k \widehat{\mathbf{g}}_k \|^2=\frac{1}{N}
\| {\bf P^\perp_{\widehat{A}_k}x} \|^2,
\end{equation}
where $\widehat{\mathbf{g}}_k=(\widehat{\mathbf{A}}_k^\dag\widehat{\mathbf{A}}_k)^{-1}
\widehat{\mathbf{A}}_k^\dag \mathbf{x}$.} Under $H_0$, 
the problem at hand is
\begin{align}
	&\max_{\sigma^2}\biggl[-N\log(\pi)-N\log(\sigma^2)-\frac{1}{\sigma^2}\|\mathbf{x}\|^2\biggr]
    \nonumber
    \\
    &=C-N\log(\bf x^\dag x).
	\label{eq_last_comprH0}
\end{align}
Gathering the above results, the final expression for the detector (up to irrelevant constants) is
\begin{equation}
	\label{eqn:penLRT01}
	\max_{k\in\{1,2,\dots,K_{\max}\}} \Biggl\{ 
    N\log \left[\frac{\bf x^\dag x}{\bf x^\dag P^\perp_{A_k} x}\right]
    -\frac{6k}{2} (1+\rho) \Biggr\}\testest \eta
\end{equation}
and it will be referred to in the following as  KLIC-based
Detector (KLIC-D).

\subsection{Estimation procedure of the $\mathbf{A}_k$s based on compressive sensing}
\label{sec_compr}

To effectively apply the compressive sensing paradigm, it is necessary to bring to light the sparse nature of the signal model. To this end, we consider an augmented steering matrix, say $\mathbf{A}$, which includes all the possible values assumed by $\mathbf{p}_k$ and a corresponding vector $\mathbf{g}$, where the nonzero entries are those associated with the $\mathbf{p}_k$s that are true. 
As a consequence, $\mathbf{g}$ is sparse, $\mathbf{A}$ represents the dictionary of the compressed sensing algorithm, and the signal model becomes
\begin{equation}
    \mathbf{x}= \mathbf{A} \mathbf{g} + \mathbf{w},
    \label{alternative_equation_model}
\end{equation}
where $\mathbf{A} = [ \mathbf{a}(\mathbf{p}_1), ... ,  \mathbf{a}(\mathbf{p}_{K_p})] \in\mathbb{C}^{N \times K_p}$ and $\mathbf{g} \in\mathbb{C}^{K_p \times 1}$, with $K_p \gg K_{\max}$ to account for $K_{\max}$ alternative hypotheses and to guarantee sparsity.

{\color{black} At the design stage, we assume that $\mathbf{g}$ obeys a sparsity-promoting prior. Actually, in literature, several sparsity-promoting priors can be found that exhibit
different properties. However, in what follows, we adopt a Laplace-based improper prior
(i.e., it is not guaranteed that its integral is finite) 
\cite{murphy2012machine,9904313} since it allows for mathematical
tractability (in fact, we come up with a closed-form expression for the estimate of the prior
parameter) and, even though it does not share any known optimality property, reliable estimation
results can be obtained from the application of this prior 
(or suitable modifications of it). Specifically, its expression is given by}
\begin{equation}
    g(\mathbf{g}; \alpha) = \frac{\alpha^{K_p}}{(2\pi)^{K_p}} \exp \biggl\{-\sqrt{\alpha} \biggl(\sum\limits_{k=1}^{K_p} \mid g_k \mid +1\biggr)\biggr\}
    \label{Laplace-based unproper prior}
\end{equation}
with $\alpha>0$ a parameter that controls sparsity. Thus, the optimization problem can be written as
\begin{align}
    \max\limits_{\mathbf{g},{\alpha}} \log \left[f_{1,K_P}(\mathbf{x}, \mathbf{g}; \sigma^2, \alpha) \right].
\end{align}
Now, notice that under $H_k$, the joint PDF of $\mathbf{x}$ and $\mathbf{g}$ can be written as
\begin{align}
    &f_{1,K_P}(\mathbf{x}, \mathbf{g}; \sigma^2,\alpha) \\ \nonumber
    &= f_{1,K_P}(\mathbf{x} \mid \mathbf{g}; \sigma^2) g(\mathbf{g}; \alpha)\\ \nonumber
    &=\frac{1}{(\pi)^{N}(\sigma^2)^N} \exp \biggl\{-\frac{1}{\sigma^2} \|\mathbf{x-A}\mathbf{g}\|^2\biggr\} \\ \nonumber
    & \times \frac{\alpha^{K_p}}{(2\pi)^{K_p}}  \exp \biggl\{-\sqrt{\alpha} \biggl(\sum\limits_{k=1}^{K_p} \mid g_k \mid +1\biggr)\biggr\}.
\end{align}
Then, taking the logarithm, the optimization problem can be recast as
\begin{equation}\begin{split}
        &\max_{\alpha}\max_{\mathbf{g}} \biggl[-N\log(\pi)-N\log(\sigma^2)-\frac{1}{\sigma^2}\|\mathbf{x - Ag}\|^2  \\
        &+K_p\log(\alpha)-K_p\log(2\pi) -\sqrt{\alpha}\biggl(\sum\limits_{k=1}^{K_p} \mid g_k \mid +1\biggr)\biggr].\end{split}
    \label{maximization_equation}
\end{equation}

\begin{figure*}
    \centering
    \includegraphics[width=0.9\linewidth]{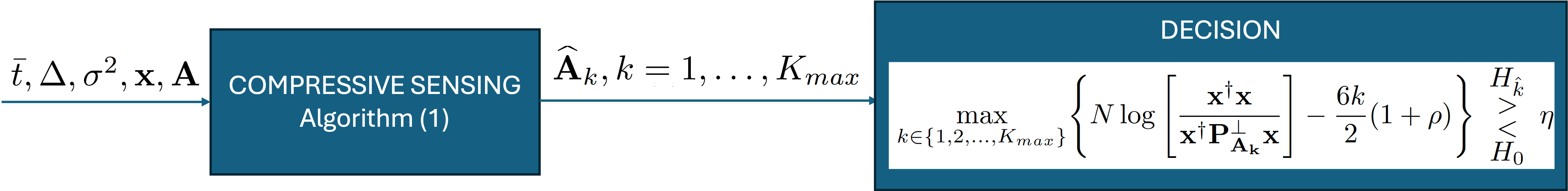}
    \caption{Block scheme for the proposed multiple hypothesis test.}
    \label{fig:blockscheme}
\end{figure*}

Let us start from the optimization with respect to $\alpha$ and notice that
\begin{equation}
    \lim_{\alpha \to 0} \log(f_{1,K_P}) = -\infty 
\label{limiteq1}
\end{equation}
and
\begin{equation}
    \lim_{\alpha \to +\infty} \log(f_{1,K_P}) = -\infty.
\label{limiteq2}
\end{equation}
Thus, setting to zero the first derivative of the objective function with respect to $\alpha$, we obtain the following equality
\begin{equation}\begin{split}
        \frac{K_p}{\alpha}-\frac{1}{2\sqrt{\alpha}}
        \biggl(\sum\limits_{k=1}^{K_p} \mid g_k \mid +1\biggr)=0,
\end{split}\end{equation}
and, hence,
\begin{equation}
    \widehat{\alpha}=\frac{4K_p^2}{\left(\sum\limits_{k=1}^{K_p} |g_k|  +1\right)^2}.
    \label{eq_alpha}
\end{equation}
Replacing \eqref{eq_alpha} in \eqref{maximization_equation}, we come up with
\begin{equation}\begin{split}
        &\max_{\mathbf{g}}\biggl[-N\log(\pi)-N\log(\sigma^2)-\frac{1}{\sigma^2}\|\mathbf{x-Ag}\|^2\\
        &+2K_p\log\Bigg(\frac{2K_p}{\sum\limits_{k=1}^{K_p} \mid g_k \mid +1}\Bigg)-K_p\log(2\pi)-2K_p\biggr]\\
        &=\max_{\mathbf{g}}\biggl[-N\log(\pi)-N\log(\sigma^2)-\frac{1}{\sigma^2} \|\mathbf{x-Ag}\|^2\\
        &+2K_p\log(2K_p)-2K_p\log\biggl(\sum\limits_{k=1}^{K_p} \mid g_k \mid +1\biggr)\\
        &-K_p\log(2\pi)-2K_p\biggr] .
\end{split}\end{equation}
It is not difficult to show that when $\norm{\mathbf{g}}\to+\infty$, the last objective function tends to $-\infty$. Thus, setting its derivative with respect to $\mathbf{g}$ to zero leads to
\begin{equation}
    -\frac{1}{\sigma^2}\biggl(\mathbf{A^{\dag}Ag-A^{\dag}x}\biggr)-K_p\frac{1}{\sum\limits_{k=1}^{K_p} \mid g_k \mid +1} \mathbf{B} \mathbf{g}=\mathbf{0} \end{equation}
with
\begin{equation}  \mathbf{B}= \text{diag}\left(\frac{1}{|g_1|},\dots,\frac{1}{|g_{K_p}|}\right),
\end{equation}
and the final result is 
\begin{equation}
    {\mathbf{g}}=\biggl[\frac{1}{\sigma^2}\mathbf{A^{\dag}A}+\frac{K_p}{\sum\limits_{k=1}^{K_p} \mid g_k \mid +1}\mathbf{B}\biggr]^{-1}\frac{1}{\sigma^2}\mathbf{A^{\dag}x}.
\label{eqn:g_solution}
\end{equation}
We proceed by iterating \eqref{eqn:g_solution} as follows
\begin{align}
    \mathbf{\widehat{g}}^{(t+1)}&=\biggl[\frac{1}{\sigma^2}\mathbf{A^{\dag}A}+\frac{K_p}{\sum\limits_{k=1}^{K_p} \mid g_k^{(t)} \mid +1}\mathbf{B}^{({t})}\biggr]^{-1}\frac{1}{\sigma^2}\mathbf{A^{\dag}x}
    \\
    &=\mathbf{C}^{(t)}\mathbf{A^{\dag}}\left[
    \sigma^2 \mathbf{I}+\mathbf{A C A^{\dag}}
    \right]^{-1}\mathbf{x},
\end{align}
where the last equality derives from the Woodbury identity \cite{lutkepohl1997handbook} and 
\begin{equation}
    \mathbf{C}^{(t)}= \frac{\sum\limits_{k=1}^{K_p} \mid g_k^{(t)} \mid +1}{K_p}\text{diag}\left({|g_1^{(t)}|},\dots,{|g_{K_p}^{(t)}|}\right).
\end{equation}
Following the lead of \cite{1336828}, it is possible to show that the iteration procedure gives rise to a nondecreasing sequence of the objective function.
We stop iterating when $t$ reaches a chosen value $\bar{t}$ or when a suitable stopping criterion is satisfied as indicated in Algorithm \ref{alg:compressive} where a schematic summary of the procedure is reported, including the initialization of the $\widehat{g}_k$s. \textcolor{black}{In this respect, we consider a matched-filter-based initialization
strategy that projects data vector onto each column of the dictionary.
This kind of initialization exploits the properties of the matched filter and 
the inherent correlation
between the measurements and the dictionary. It is clear that
other strategies are possible, such as a random selection of the initial
point, but, as shown in the next sections, this matched-filter-based
strategy allows for a stable convergence of the iterative algorithm
and leads to reliable performance.} Therein, the parameter $\epsilon>0$ allows for a control of the number of iterations. The required value for $\bar{t}$ is determined by analyzing the convergence rate.
After $\bar{t}$ iterations, we select the $k<N$ highest peaks of $\mathbf{\widehat{g}}^{(\bar{t})}$ to form $\mathbf{\widehat{A}}_{k}, k=1,\dots,K_{\max}$, for all alternative hypotheses in \eqref{eq_multipleHyp}. It is worth of noting that the maximum number $\bar{t}$ of iterations is independent of the maximum number $K_{\max}$ of scatterers to be detected.
Finally, we compute the PLLRT substituting the parameters with the respective estimates as depicted in Figure \ref{fig:blockscheme}. 
\textcolor{black}{From inspection of Algorithm 1, the computational cost of the sparse estimation procedure
in terms of the usual Landau notation is 
$\mathcal{O}(K_{\max}(N^3+N^2+3K_PN+2N))\approx \mathcal{O}(K_{\max}N^3)$, where
$K_{\max}$ depends on the specific application.}

\begin{algorithm}[tb!]
    \caption{Compressive sensing algorithm to estimate the $\mathbf{A}_k$s}
    \label{alg:compressive}
    \begin{algorithmic}[1]
        \REQUIRE $\bar{t},\Delta, \sigma^2,\mathbf{x},\mathbf{A}$
        \ENSURE $\mathbf{\widehat{A}}_{k}, k=1,\dots,K_{\max}$
        \STATE Set $t=0$, $\widehat{g}_{k}^{(0)}=\left| {{\mathbf{\mathbf{a^{\dag}}(\mathbf{p}_k)}}\mathbf{x}}\right|$, $k=1,\dots,K_P$,
        \STATE Set $t=t+1$
        
        \STATE Compute $$\mathbf{C}^{(t-1)}=\frac{\sum\limits_{k=1}^{K_p} \mid g_k^{(t-1)} \mid +1}{K_p}\text{diag}\left({|g_1^{(t-1)}|},\dots,{|g_{K_p}^{(t-1)}|}\right)$$
        \STATE Compute $$\mathbf{\widehat{g}}^{(t)}=\mathbf{C}^{(t-1)}\mathbf{A^{\dag}}\left[
    \sigma^2 \mathbf{I}+\mathbf{A C A^{\dag}}
    \right]^{-1}\mathbf{x}$$
        \STATE If $\bar{t}=t+1$ or $\frac{||\mathbf{\widehat{g}}^{(t)}-\mathbf{\widehat{g}}^{(t-1)}||}{||\mathbf{\widehat{g}}^{(t)}||}<\epsilon$ go to step 6 else go to step 2
        \STATE Select the $k$ highest peaks of $\mathbf{\widehat{g}}^{(\bar{t})}$ and form $\mathbf{\widehat{A}}_{k}=[ \mathbf{a}(\mathbf{p}_1), ... ,  \mathbf{a}(\mathbf{p}_{k})]$ for all $k=1,\dots,K_{\max}$
        \STATE Return $\mathbf{\widehat{A}}_{k}, k=1,\dots,K_{\max}$
    \end{algorithmic}
\end{algorithm}

\section{Performance Analysis on Synthetic Data}
\label{sec_performance_sim}
In this section, we assess the nominal behavior of the proposed detector on simulated data in terms of false alarm rate, detection probabilities, and estimation error. Moreover, we compare these results with those returned by the most natural counterpart, i.e., the Sup-GLRT method \cite{budillon2015glrt}.
	
\begin{figure}
    \centering
    \includegraphics[width=\columnwidth]{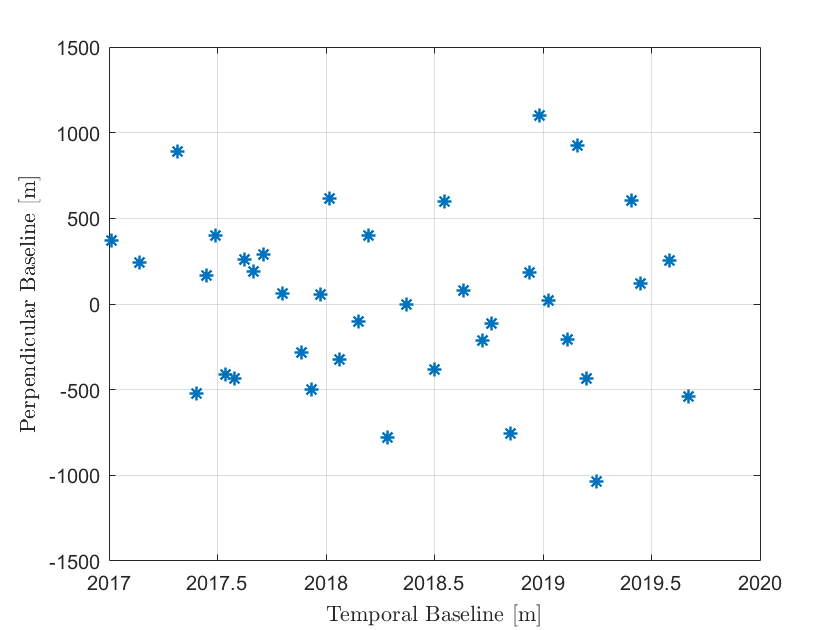}
    \caption{Distribution of the acquisitions, reported as stars, in the Temporal / Spatial Baseline domain.}
    \label{fig:plot_baseline_Centro_Direzionale}
\end{figure}

In the numerical examples, we simulated data according to the 
acquisition parameters of a real X-band data set, composed by $N=38$ SAR 
images. Particularly, the system parameters are related to the real COSMO-SkyMed dataset exploited in the next section, where the experiments are carried out on real data. The distribution of the acquisitions, in the temporal/perpendicular baseline domain, is reported in Figure 
\ref{fig:plot_baseline_Centro_Direzionale}. The spans of the perpendicular and 
temporal baselines are $\Delta B\sim2100$ m and $\Delta t=971$ 
days ($\sim$2.65 years), corresponding to nominal Rayleigh resolutions \cite{4Dfornaro} in 
elevation $\delta_s=\lambda r_0/(2\Delta B) \approx5.45$ m (related to 
the height resolution $\delta z\approx 3.08$ m, being $
\theta=34.4^\circ$) and velocity $\delta v=\lambda/(2\Delta t) \approx 
0.59$ cm/year, being $\lambda = 3.1$ cm and $r_0=745$ km.

As for the position of the scatterer in the parameter space, 
$\mathbf{p}_k$, $k=1,...,K_p$, the values of the elevation $s$ vary
in $[-177,177]$ m ($[-100,100]$ m in terms of height), while those of the velocity $v$ take on values in $[-1,1]$ cm/year. Elevation and velocity intervals are sampled with a spacing equal to half of the corresponding nominal resolution. 
{\color{black}
Notice that when the actual position of the point scatterer is not exactly
aligned with the grid points, a performance degradation can occur that depends
on the extent of the deviation between the actual steering vector and
the nominal steering vectors belonging to the search grid. In order to mitigate
this loss, an effective strategy could consist in repeating the entire estimation
procedure with the same number of grid points but the parameter 
range of variation restricted around
the estimates returned at the previous iteration. Doing so, the computational
load of each iteration would remain unaltered (namely that of the sparse
estimation procedure) while the overall computational cost
becomes proportional to the number of iterations (as shown below
the computational complexity depends on the size of the search grid
and, hence, increasing the cardinality of the grid could lead
to prohibitive execution times). As an alternative, interpolation
strategies can be pursued to reduce this straddle loss.
}
For the comparison analysis between Sup-GLRT and the KLIC-D, we assume that the maximum number of scatterers within a tomographic cell is equal to $K_{\max}=2$. Subsequently, we extend the performance analysis of KLIC-D to the case of $K_{\max}=3$.

In what follows, the figures of merit are:
\begin{itemize}
\item the Probability of Detection under $H_{i}$, say $P_{d,H_i}=P(\text{accept }\bar{H_0}|H_{i})$ as a function of the SNR $={g^2}/{\sigma^2}$, with $g>0$ the modulus of the reference PS backscattering coefficient and $\sigma^2=1$;
\item the Root Mean Square Error (RMSE) for the estimate of the number of scatterers as a function of the SNR under $H_{i}$, defined as 
$$R_{K,H_{i}} = \sqrt{\biggl(\sum\limits_{l=1}^{N_{MC}} \abs{K - \widehat{K}_i(l)}^{2}\biggr)/N_{MC}}$$
where $\widehat{K}_i(l)$ is the estimate of the number of scatterers at the $l$th Monte Carlo (MC) trial under $H_i$, while $N_{MC}$ represents the total number of MC trials;
{\color{black}\item  the probability of correct classification, say $P_{c,H_i}=P(H_i|H_i)$, $i=1,\dots,K_{\max}$;
\item the RMSE of the estimated scatterer localization in terms of height and velocity:
\[
R_{h,H_{i}} = \sqrt{\biggl(\sum\limits_{l=1}^{N_{MC}} \abs{z - \widehat{z}_i(l)}^{2}\biggr)/N_{MC}},
\]
where $\widehat{z}_i(l)$ is the estimate of scatterer height [m] at the $l$th MC under $H_i$, $i=1,\dots,K_{\max}$, and,
\[
R_{v,H_{i}} = \sqrt{\biggl(\sum\limits_{l=1}^{N_{MC}} \abs{v - \widehat{v}_i(l)}^{2}\biggr)/N_{MC}},
\]
where $\widehat{v}_i(l)$ is the estimate of scatterer velocity [cm/year] at the $l$th MC under $H_i$, $i=1,\dots,K_{\max}$.}
\end{itemize}
The above estimates are obtained over 5000 independent MC trials, while the detection thresholds are computed via $100/P_{fa}$ runs, with $P_{fa} = 10^{-3}$.
The value of the parameter $\rho$, in \eqref{eq_penalty}, is set to 3, since this choice ensures a probability of selecting hypothesis $H_2$ under $H_1$, i.e., $P(\text{accept }H_2|H_1)$, equal to $10^{-3}$ for $K_{\max}=2$ and $\text{SNR}=15$ dB. \textcolor{black}{For $K_{\max}=3$, the parameter $\rho$ is set to 5 in order to guarantee the same misclassification probability.}

\subsubsection{Convergence analysis and False Alarm probability}
\begin{figure}[tpb!]
		\begin{center}
\includegraphics[width=\columnwidth]{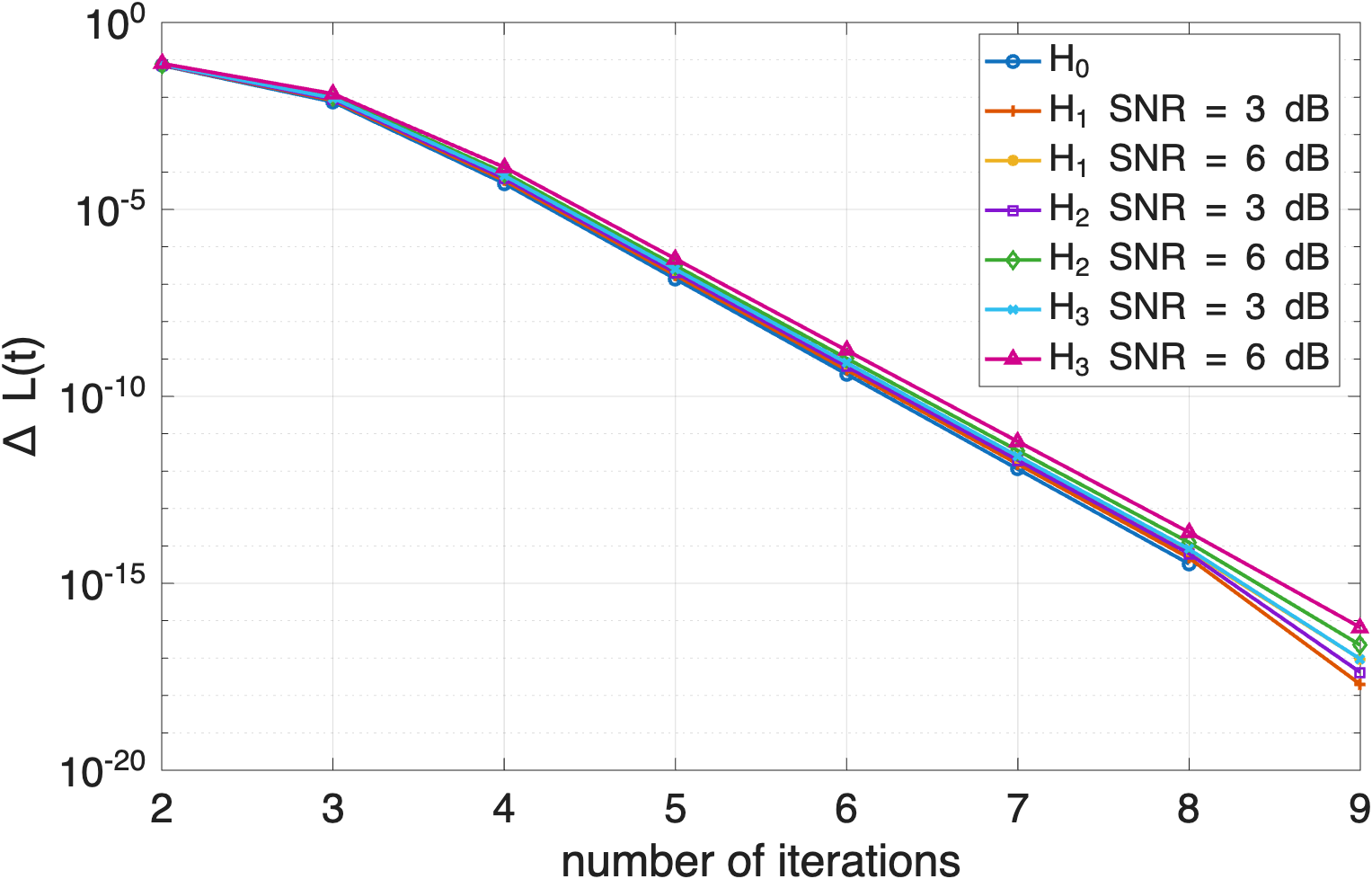}
			\caption{\textcolor{black}{Relative variation of the log-likelihood   \eqref{maximization_equation} versus the number of iterations for the compressive sensing algorithm.} }
			\label{fig:figure01}
		\end{center}
	\end{figure}

As a preliminary step, we analyze the necessary number of iterations $\bar{t}$ required for the compressive sensing algorithm and its convergence rate. To this end, in Figure \ref{fig:figure01}, we plot the values of the relative variation of the compressed log-likelihood, i.e., 

 	\begin{equation}
		\Delta L(t) = \abs{\biggl( L(t)-L(t-1)\biggr)/L(t)}
	\end{equation}
 
where
 
 	\begin{equation}
		L(t)=\log \left[f_{1,K_P}(\mathbf{x}, \widehat{\mathbf{g}}^{(t)}; \sigma^2, \widehat{\alpha}) \right],
	\end{equation}
with $\sigma^2=1$, averaged over 1000 MC versus $t$, under $ H_{0} $, $H_{1}$, $H_2$, and $H_3$  with $\text{SNR}= 3,6$ dB. It turns out that, for the chosen parameters, 6 iterations are sufficient to achieve a variation lower than $10^{-8}$ for all the situations considered. Thus, in what follows, we set $\bar{t} = 6$.

\begin{figure}[tbp!]
		\begin{center}
\includegraphics[width=\columnwidth]{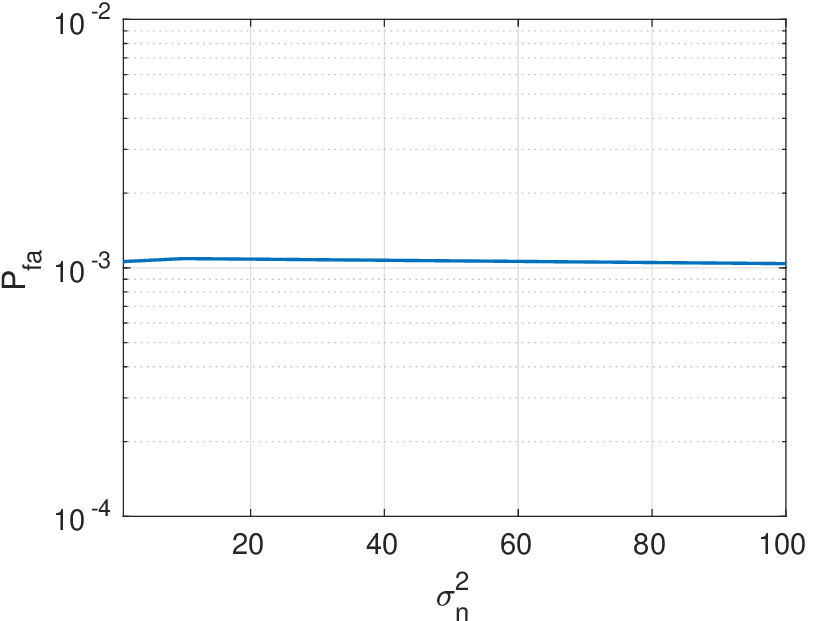}
			\caption{Sensitivity analysis of $P_{fa}$ a with respect to mismatched $\sigma^2$ parameter (threshold is estimated with $\sigma^2=1$).}
			\label{fig:CFAR}
		\end{center}
	\end{figure}

Moreover, we evaluate the robustness of the PFA in the presence of a mismatch between the assumed and actual noise variance. Particularly,
Figure \ref{fig:CFAR} presents a sensitivity analysis of $P_{fa}$ as a function of the noise variance parameter  which ranges from 1 to 1000, whereas the threshold is estimated by assuming $\sigma^2=1$.
The analysis reveals that as the noise variance increases well beyond the value used to estimate the nominal detection threshold, the PFA remains constant, at least for the considered simulation setup. For this reason, in the following and in the real data analysis, we set $\sigma^2=1$.

\subsubsection{Comparison with the Sup-GLRT}
Here, we compare the performance of the KLIC-D with  the Sup-GLRT proposed in \cite{budillon2015glrt}. 
For the ease of the reader, we provide the expression of the Sup-GLRT:
the decision at the $k$th stage, with $k=1,\dots,K_{\max}$, compares the statistic to a threshold to select  either $H_{k-1}$ or $H_{K_{\max}}$, that is
\begin{equation}
\Lambda_k(\mathbf{x}) = \frac{\min\limits_{\mathbf{A}_{k-1}} \mathbf{x}^{\dag} \mathbf{P}^\perp_{\mathbf{A}_{k-1}} \mathbf{x}}{\min\limits_{\mathbf{A}_{K_{\max}}} \mathbf{x}^{\dag} \mathbf{P}^\perp_{\mathbf{A}_{K_{\max}}} \mathbf{x}}  \test \eta_k,
    \label{Sup-GLRT}
\end{equation}
where $\eta_k$, $k=1,\dots,K_{\max}$, are the detection thresholds.
It should be noted that \eqref{Sup-GLRT} requires a sequential multistage detection process, where each stage requires a thresholds. On the other hand, KLIC-D exploits a single adaptive threshold for all hypotheses (in addition to a tuning parameter). This significantly simplifies threshold setting. 
To limit the computational burden of Sup-GLRT, in the following analysis, we set $K_{\max}=2$.

\begin{figure}[htp!]
		\begin{center}
			\includegraphics[width=\columnwidth]{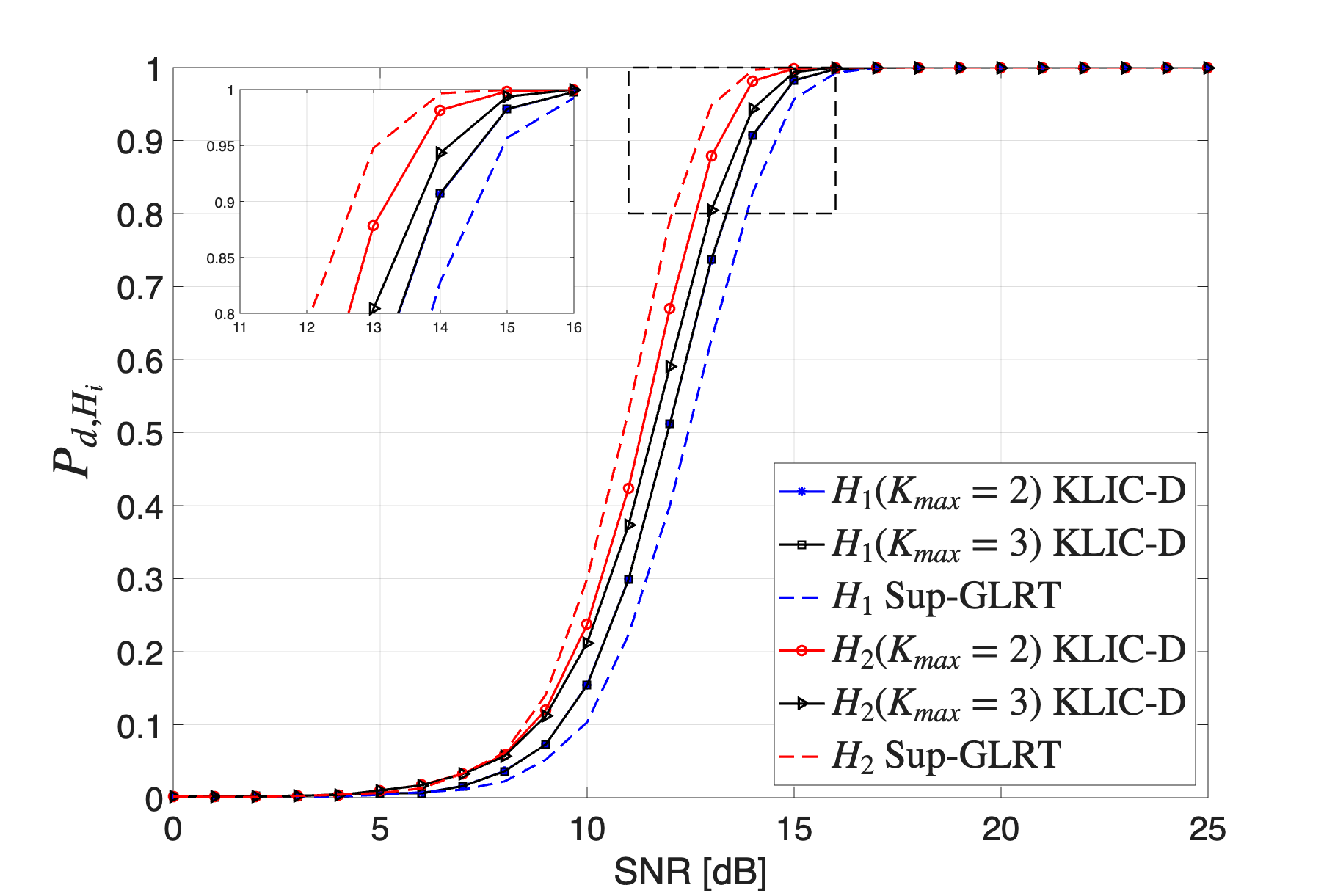}
            \includegraphics[width=\columnwidth]{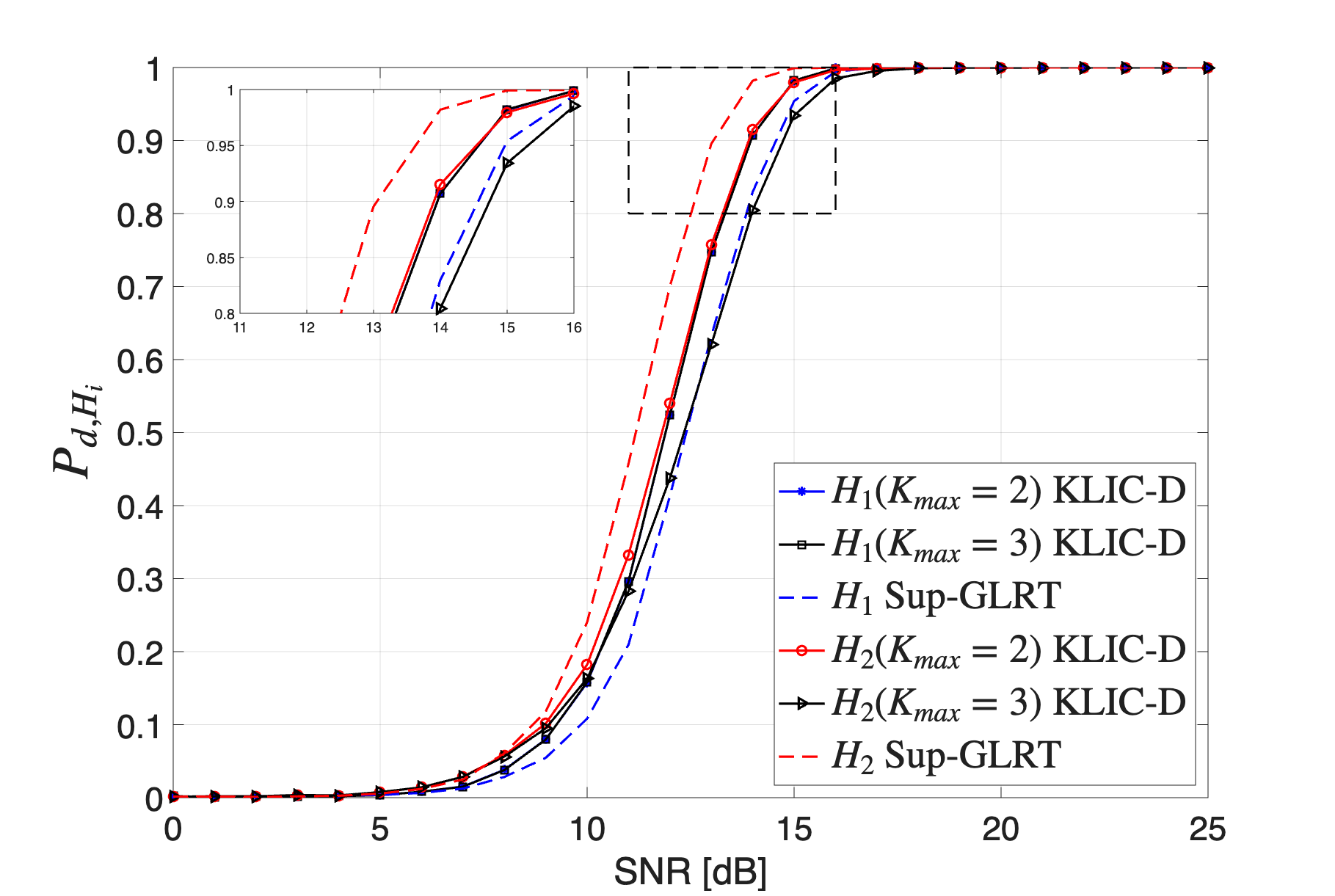}
			\includegraphics[width=\columnwidth]{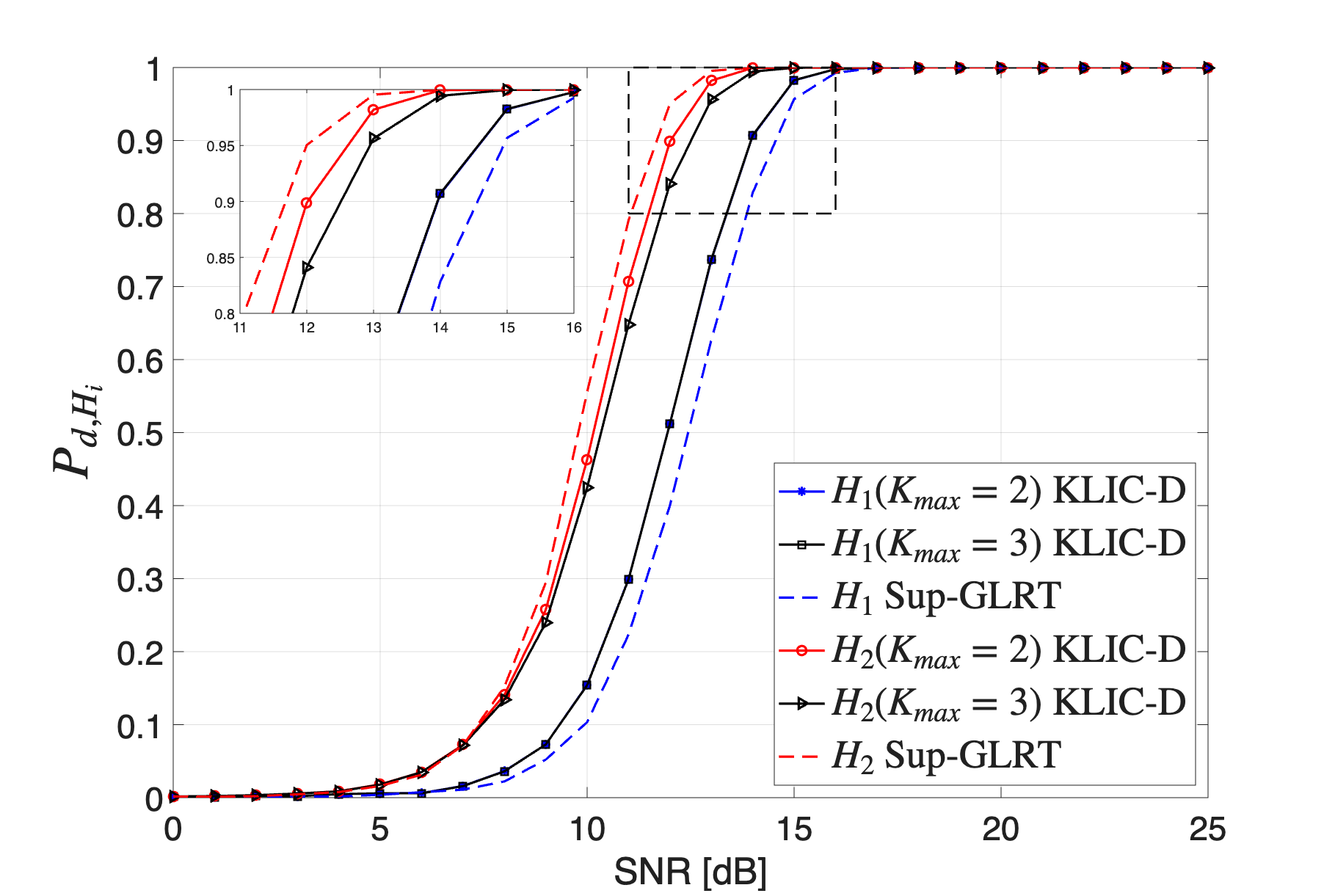}
            \caption{{Probability of Detection $P_{d,H_i}$, $i=1,2$  versus SNR [dB], assuming the same scattering coefficients and fixed positions (at the top) and random positions (at the middle), assuming different scattering coefficients and fixed positions (at the bottom).}}
			\label{fig:P_d_comp}
		\end{center}
\end{figure}

\begin{figure}[htp!]
		\begin{center}
			\includegraphics[width=\columnwidth]{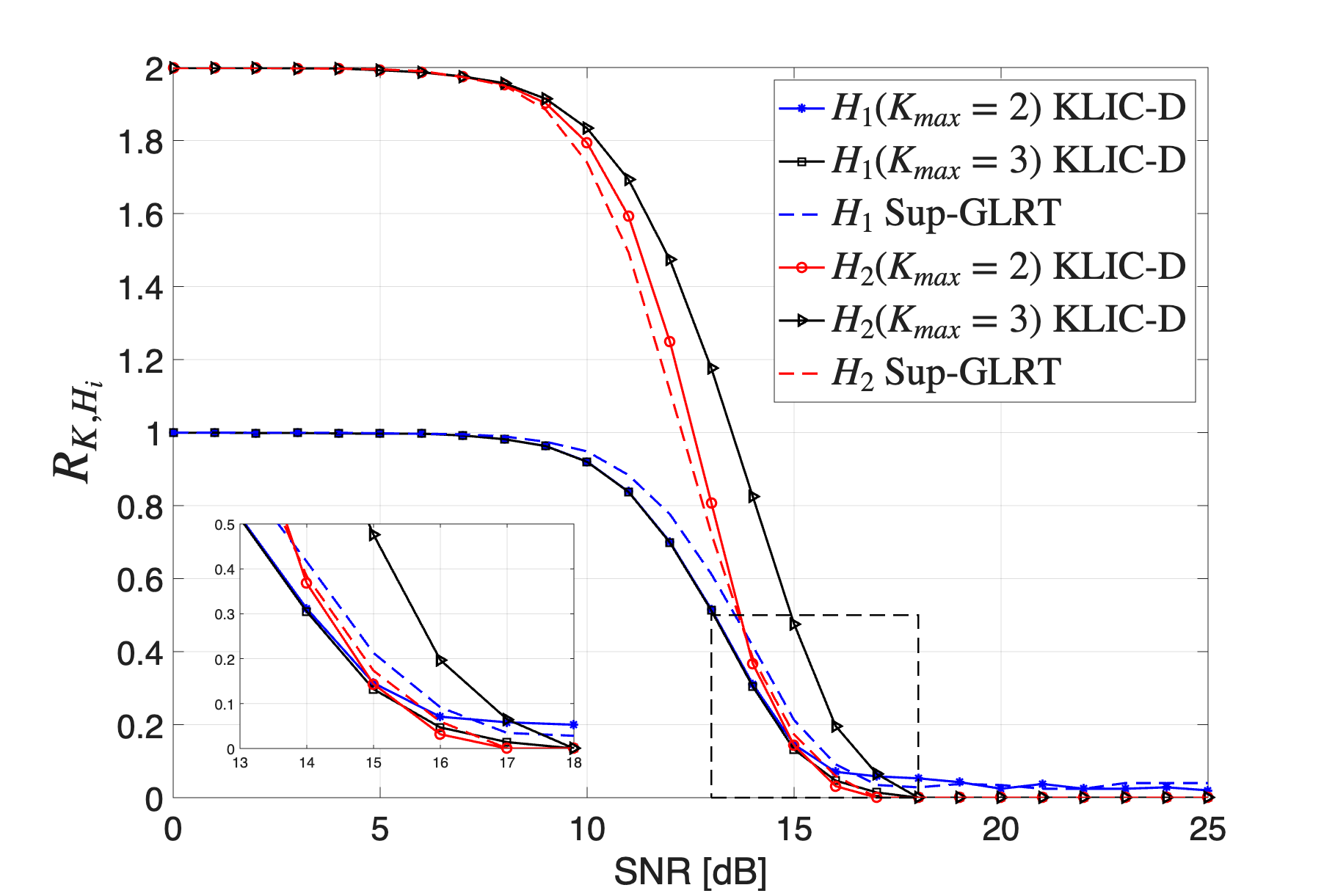}
            \includegraphics[width=\columnwidth]{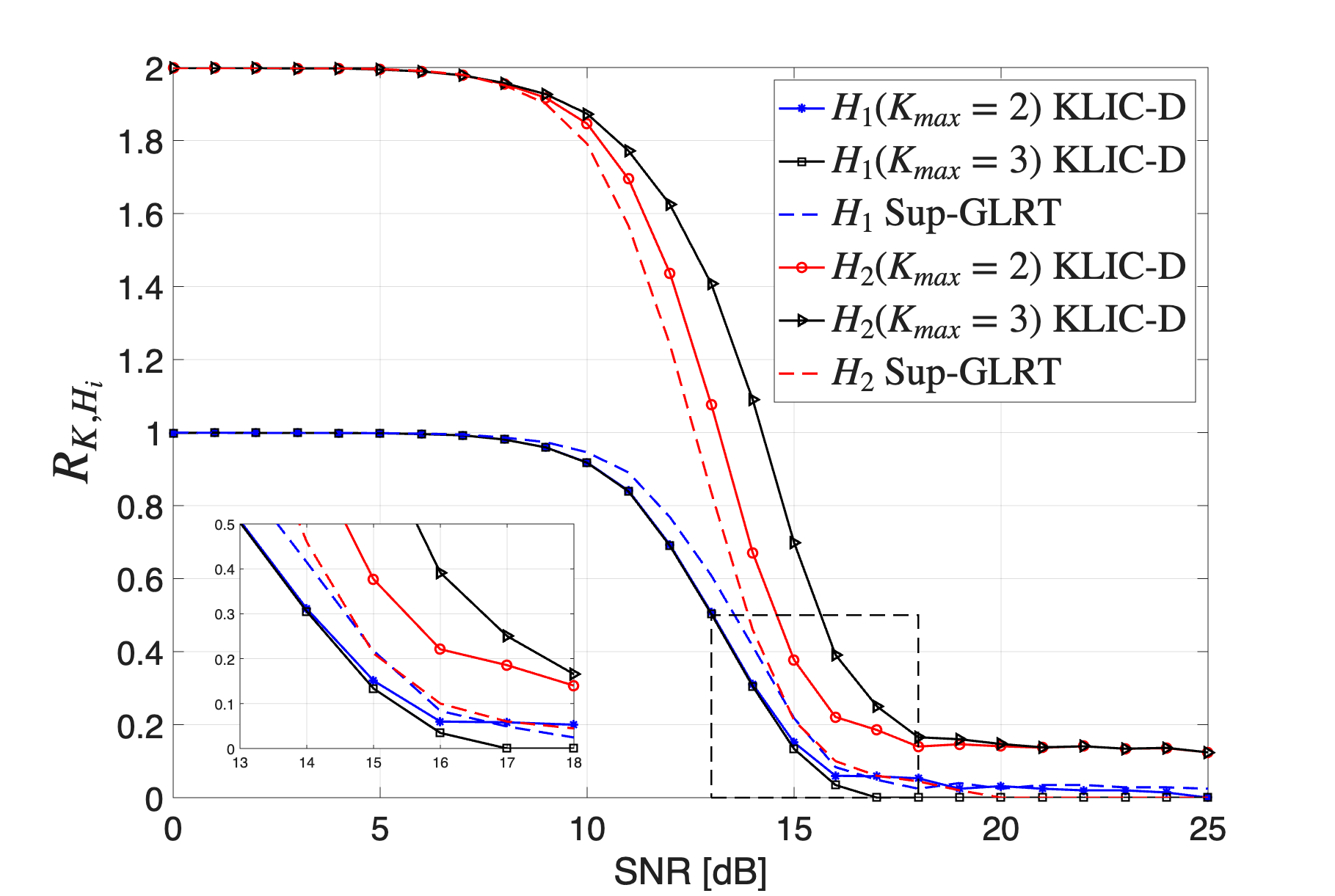}
			\includegraphics[width=\columnwidth]{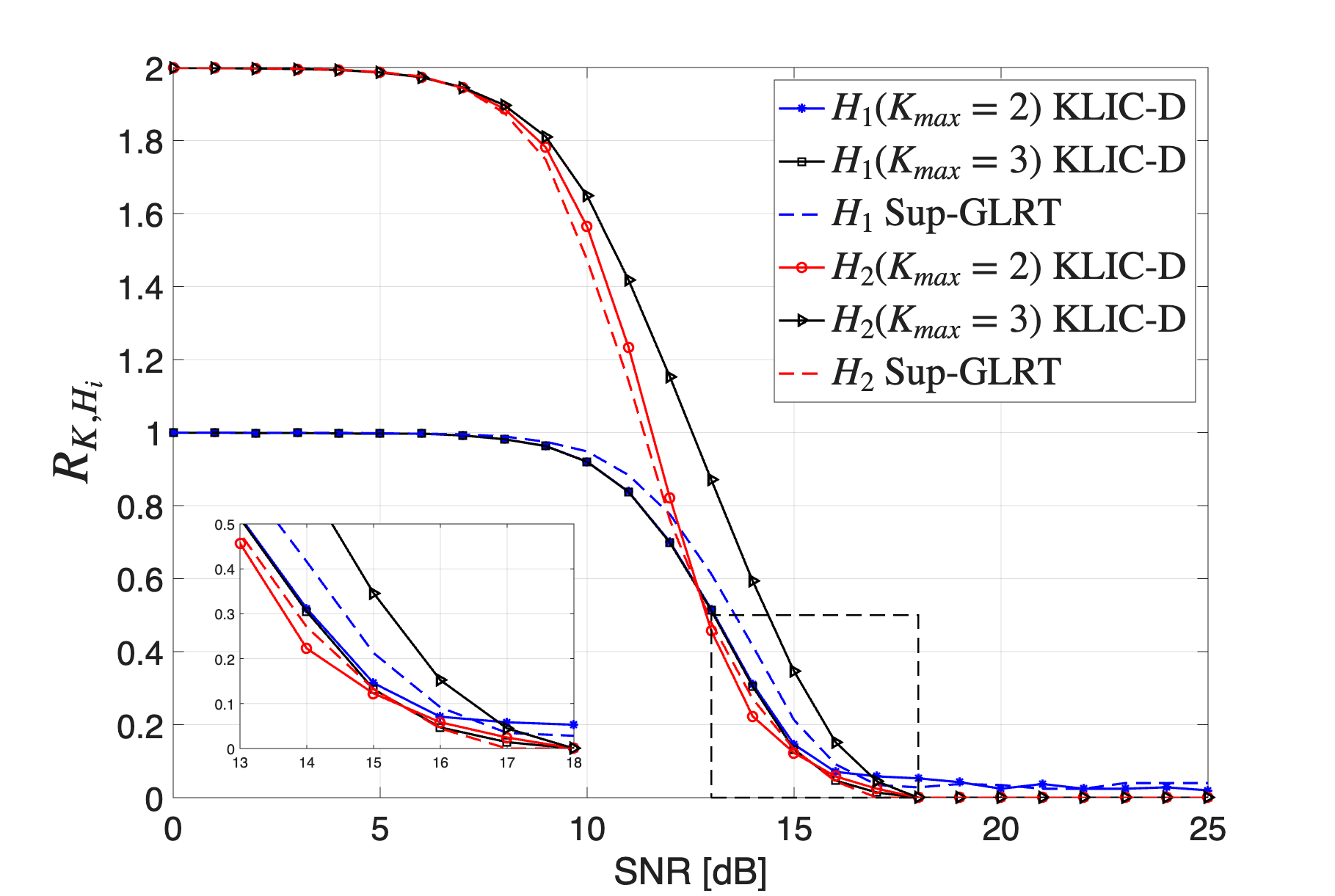}			\caption{{Root Mean Square Error $R_{K,H_i}, i=1,2$ versus SNR [dB], assuming the same scattering coefficients and fixed positions (at the top) and random positions (at the middle), assuming different scattering coefficients and fixed positions (at the bottom).}}
			\label{fig:RMS_comp}
		\end{center}
\end{figure}

Figures \ref{fig:P_d_comp} and \ref{fig:RMS_comp} show the results of the comparison.
In particular, Figure \ref{fig:P_d_comp} show the results in terms of $P_{d,H_i}, i=1,2$, as functions of the SNR that takes on values in the interval $[0,25]$ dB, when $H_{i}, i=1,2$ are in force (the true hypothesis is indicated in the figure's legend). We consider two different configurations for the proposed detector: $K_{\max}=2$ and $K_{\max}=3$ (please notice that it is not easy to consider the latter case for Sup-GLRT from the computational point of view, as the implementation of the projector into the subspace spanned by all possible combinations of the three scatterers is required in \eqref{Sup-GLRT}).
The analysis is carried out considering different configurations in terms of scatterer position and power.  
In the first case of Figure \ref{fig:P_d_comp} (top), the scatterers have the same power levels and fixed positions (the first target is located in the center of the steering matrix, while the second target is placed at an elevation distance of 30.8 m and with the same velocity). The detection probability under $H_2$ is higher than under $H_1$, since multiple scatterers contribute to increase the backscattered power within the same tomographic cell. The detectors share a similar detection performance under both hypotheses. In the second case (middle subfigure), the scatterers are randomly positioned within the nominal resolution cells. Performance slightly degrades compared to the fixed-position scenario due to the spatial variability.
In the third case (bottom), the scatterers exhibit different 
power levels:\footnote{{\color{black} For simplicity, we assume that the phases of the $g_k$s are equal to zero.}}
\begin{itemize}
\item under $H_1$, the backscattering coefficients are set as $|g_1|^2={ g^2}$;
\item under $H_2$, they are defined as $|g_1|^2={g^2}$ and $|g_2|^2={1.5  g^2}$, leading to an increased power imbalance.
\end{itemize}
As expected, the detection probability when $H_2$ is in force is better than under $H_1$ since the additional scatterer has higher power. The detection probability benefits from the presence of multiple scatterers, especially at higher SNR values.

Figure \ref{fig:RMS_comp} illustrates $R_{K,H_{i}}$ as a function of SNR. As expected, at low SNR, the estimation error is relatively high because weaker scatterers are harder to distinguish from noise.
As SNR increases, the RMSE significantly decreases, indicating a more accurate estimation of the number of scatterers. When scatterers have different power levels (bottom subfigure), the RMSE is slightly higher than in the equal-power cases. This is expected, as the imbalance in backscattering power can lead to a wrong estimation of the scatterer number, particularly for high SNR. 

{\color{black}
Figure \ref{fig:P_{c,H_i}_comp} illustrates the probability of correct classification versus SNR for the proposed KLIC-D and the Sup-GLRT under different scatterer configurations. As expected, at low SNR, this probability is relatively low, but, as the SNR increases, it significantly increases under both hypotheses, indicating a more accurate estimation of the number of scatterers.
In scenarios with equal scattering coefficients and fixed positions (top subfigure), KLIC-D achieves performance comparable to the competitor, successfully leveraging the increased backscattered power from multiple scatterers to enhance $P_{c,H_2}$. 
While KLIC-D remains computationally feasible and structurally scalable to $K_{\max}=3$, the probability of correct classification is lower than that in scenarios with $K_{\max}=2$. This would be a natural behavior as the increased number of alternative hypotheses introduces additional uncertainty in distinguishing the exact number of scatterers, particularly at low to moderate SNR levels. 
The introduction of random scatterer positions leads to a modest performance degradation due to spatial variability, yet the detector maintains reliable performance. In the cases with unbalanced scattering coefficients, KLIC-D obtains better performance, particularly at moderate to high SNR levels.

Figures \ref{fig:newRMSEH1}-\ref{fig:newRMSEH2b} show the RMSE for estimating the height and velocity under $H_1$ and $H_2$. The monotonic decrease in RMSE with increasing SNR for both methods confirms the expected estimation behavior. Notably, the proposed KLIC-D exhibits localization accuracy that is almost identical to or better than the Sup-GLRT  across the entire SNR range. This result validates that the integration of the compressive sensing paradigm within the KLIC-D framework does not compromise parameter estimation quality. 
For the fixed-position configurations, varying the scatterer powers did not significantly affect height RMSE, likely due to the dominance of the spatial separation in the estimation error.

Overall, these results validate the effectiveness of the proposed
method in detecting multiple scatterers even under moderate
SNR conditions. The results also demonstrate robustness against power imbalances while maintaining comparable
detection performance with respect to Sup-GLRT. Moreover, while the Sup-GLRT is limited in practice to $K_{\max} = 2$ due
to its sequential, multi-stage detection process, the proposed method can be set to manage more than two scatterers, while
maintaining reliable performance in the presence of two scatterers.
This is corroborated by the performance analysis, where in the presence of $2$ PSs, the detector maintains a high probability of
detection and correct classification as well as low 
RMSE values even with $K_{\max} = 3$.}
{\color{black}
In some cases, for high SNR values, the RMSE returned by the proposed procedure could be slightly higher than that of the Sup-GLRT. This behavior might be due to the fact that for high SNR values the sidelobes of the sparse response associated with each scatterer  might interfere with the respective mainlobes.
}

\begin{figure}[htp!]
		\begin{center}
	       \includegraphics[width=\columnwidth]{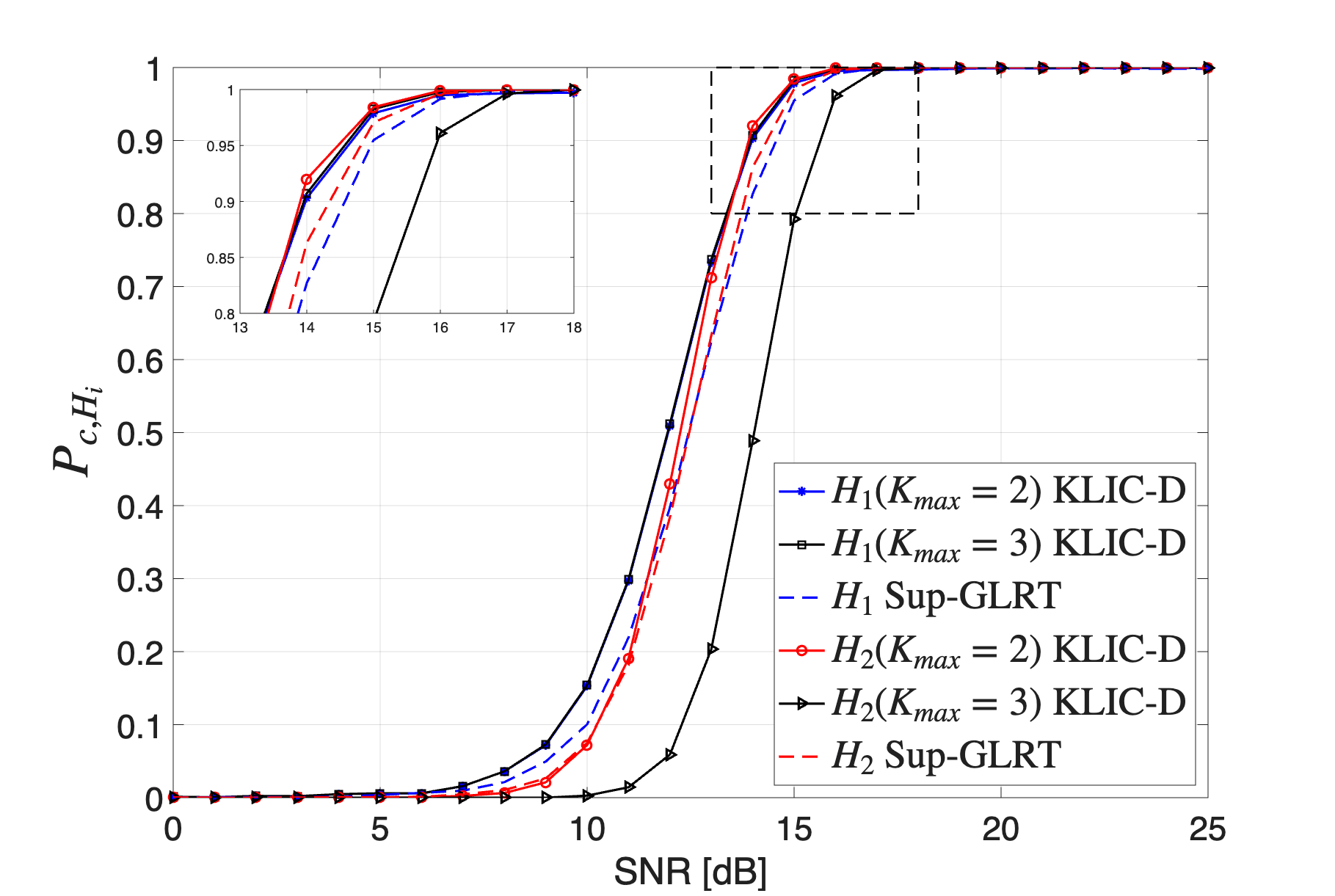}
            \includegraphics[width=\columnwidth]{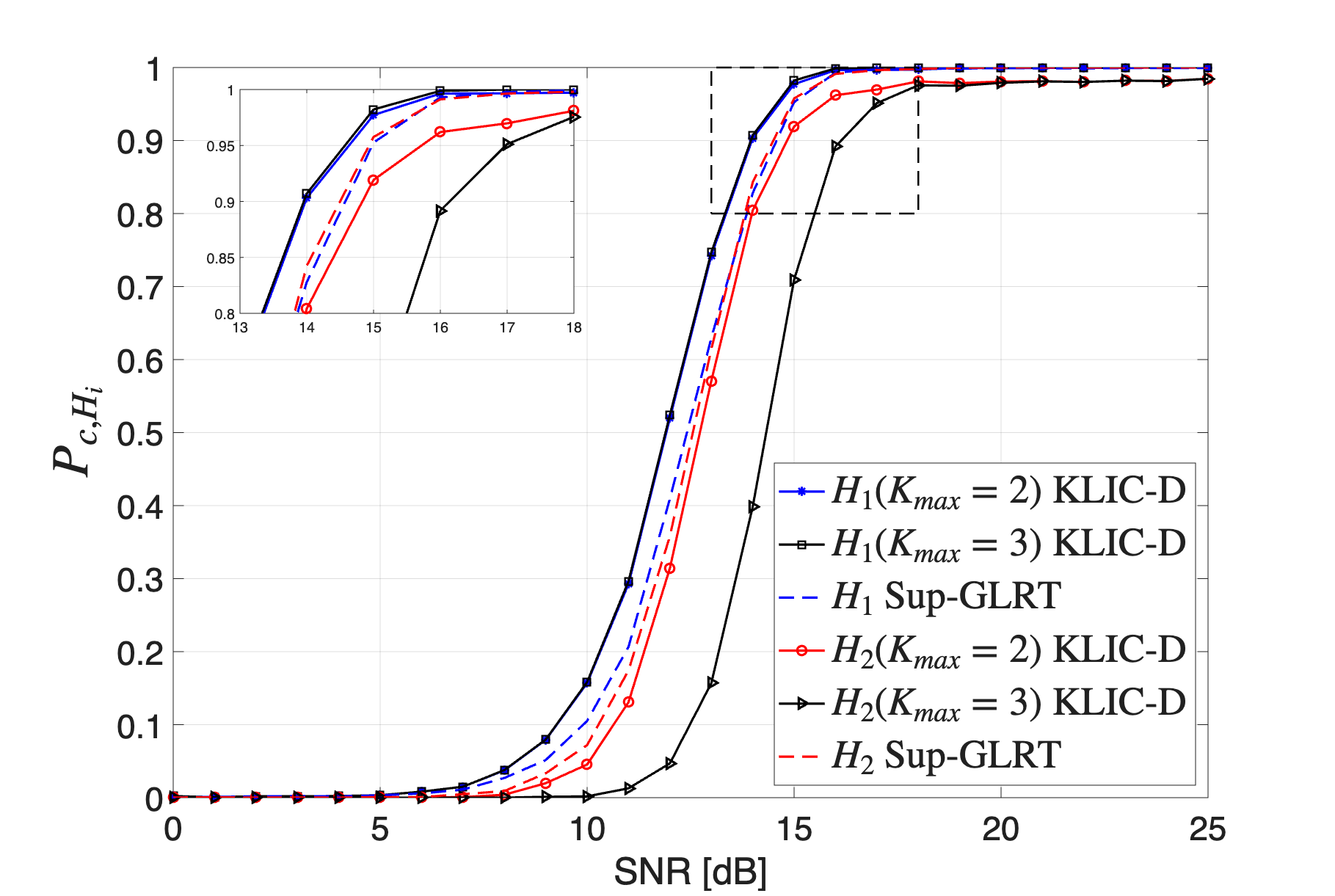}
			\includegraphics[width=\columnwidth]{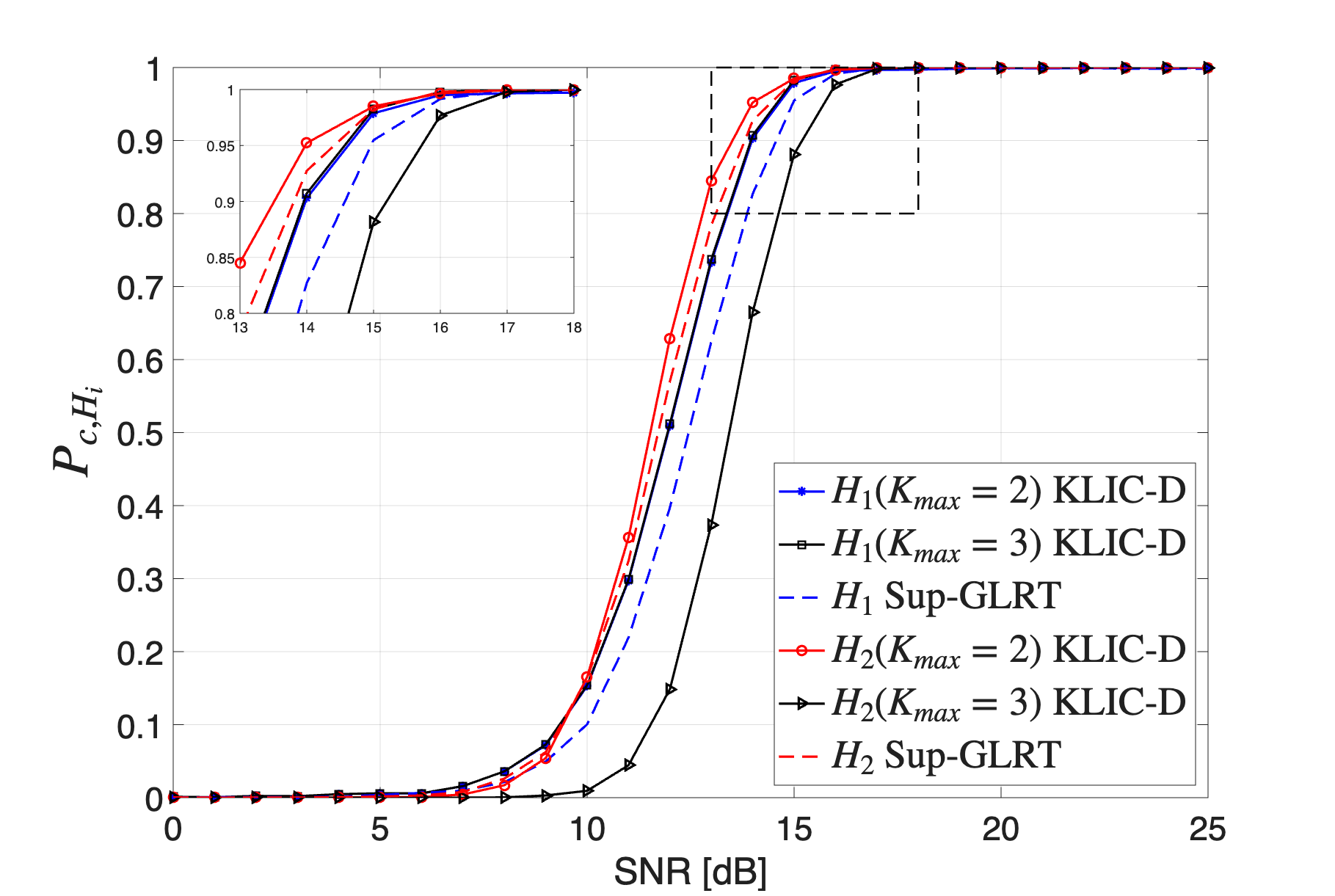}
            \caption{Probability of correct classification $P_{c,H_i}$  versus SNR [dB], assuming the same scattering coefficients with fixed positions (at the top), random positions (at the middle), and assuming different scattering coefficients with fixed positions (at the bottom).}
			\label{fig:P_{c,H_i}_comp}
		\end{center}
\end{figure}

\begin{figure}
    \centering
\includegraphics[width=\linewidth]{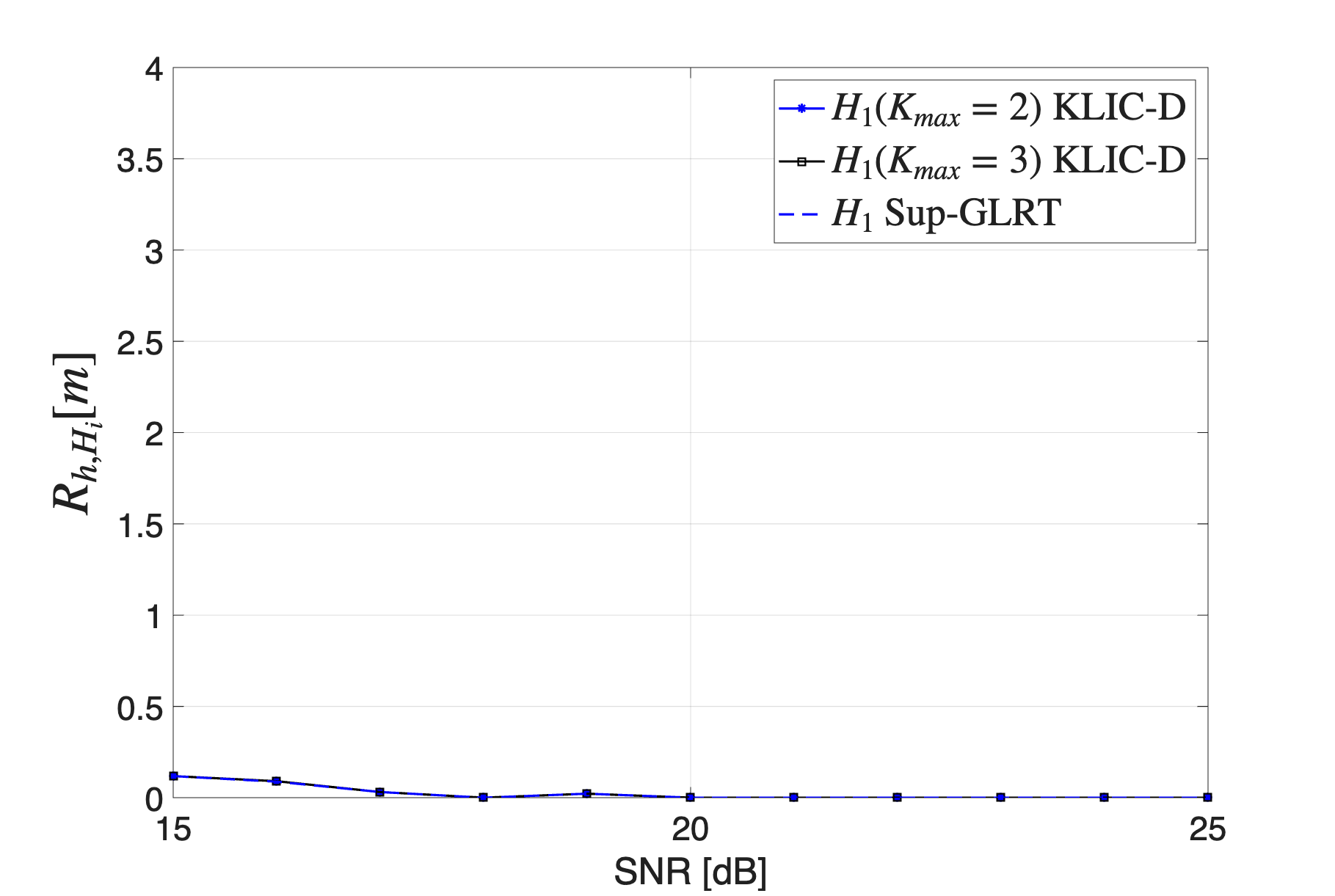}
\includegraphics[width=\linewidth]{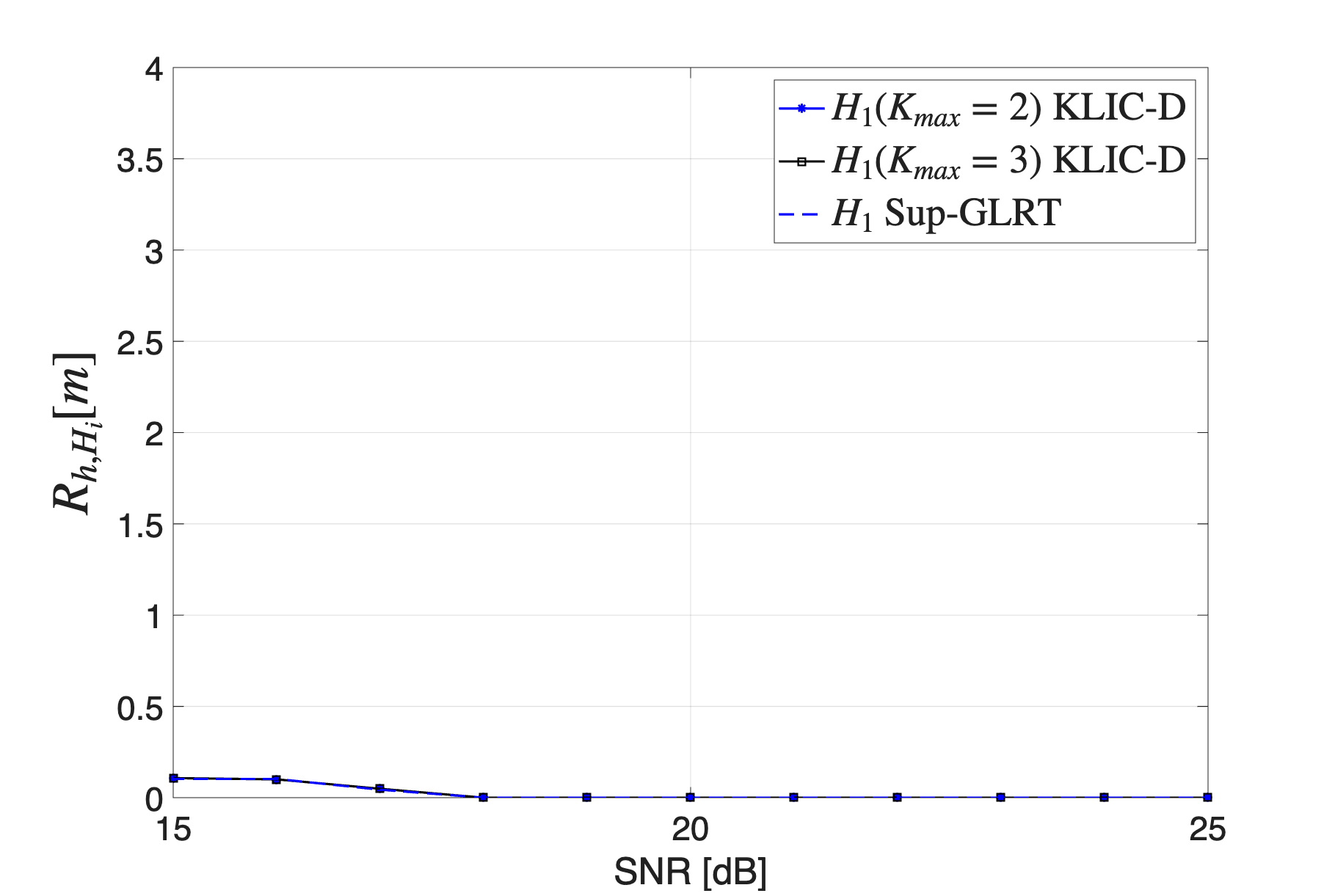}
\caption{RMSE for scatterer height estimation [m] versus SNR under $H_1$, assuming fixed positions (at the top) and random positions (at the  bottom).} 
    \label{fig:newRMSEH1}
\end{figure}

\begin{figure}[tbp]
    \centering
\includegraphics[width=\linewidth]{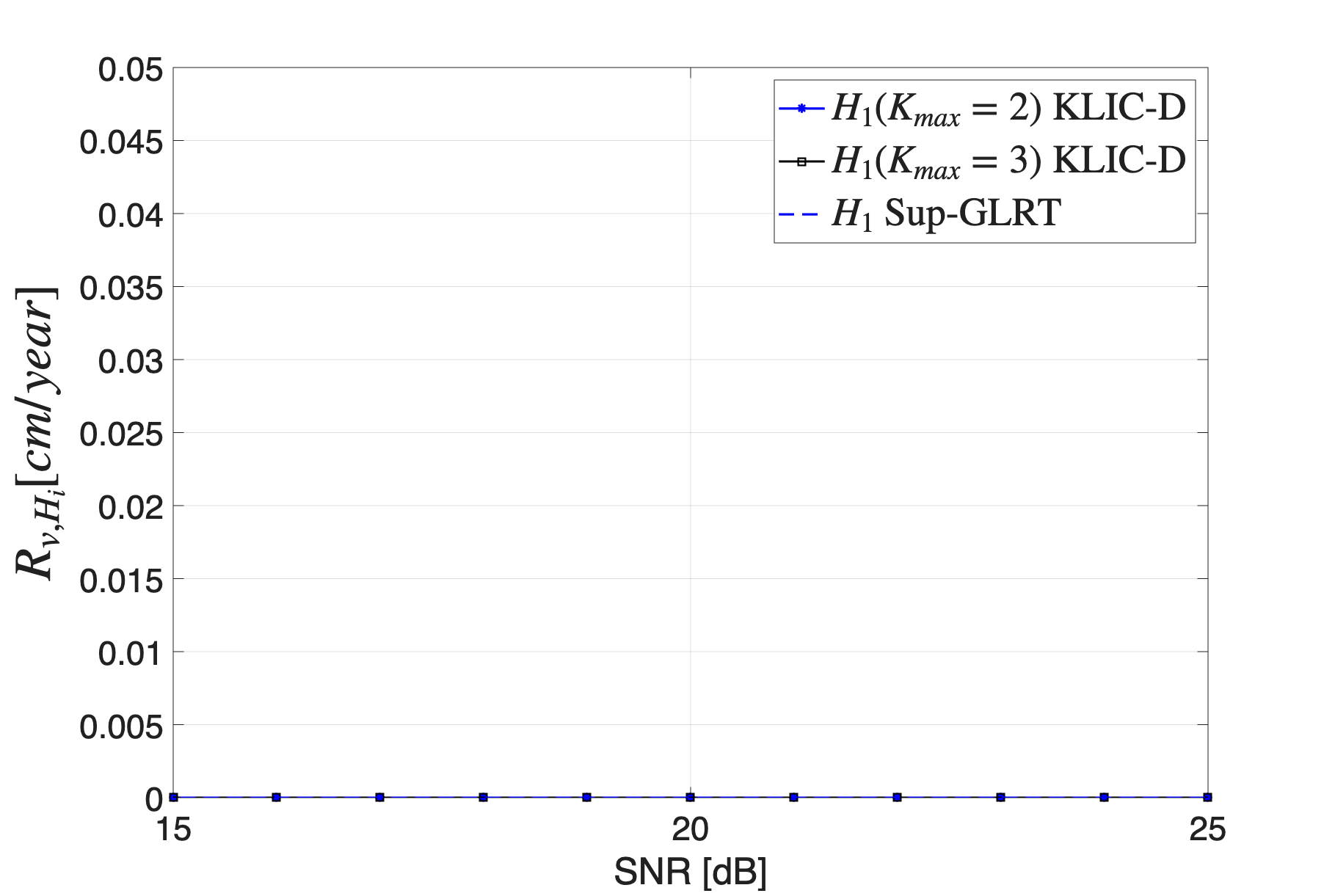}
\includegraphics[width=\linewidth]{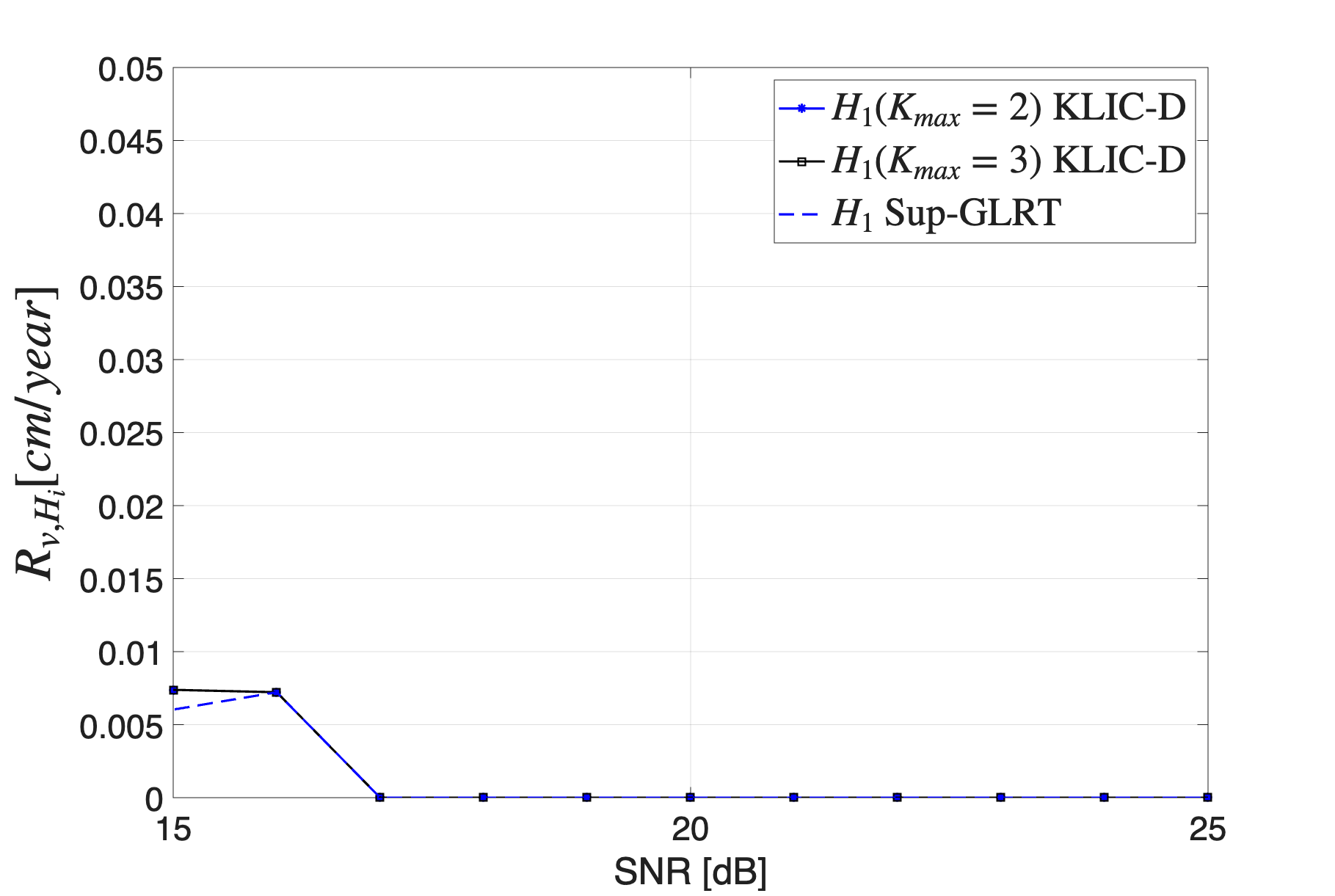}
\caption{RMSE of scatterer velocity estimation [cm/year] versus SNR under $H_1$, assuming fixed positions (at the top) and random positions (at the bottom). }
    \label{fig:newRMSEH1b}
\end{figure}

\begin{figure}[tbp]
    \centering
\includegraphics[width=\linewidth]{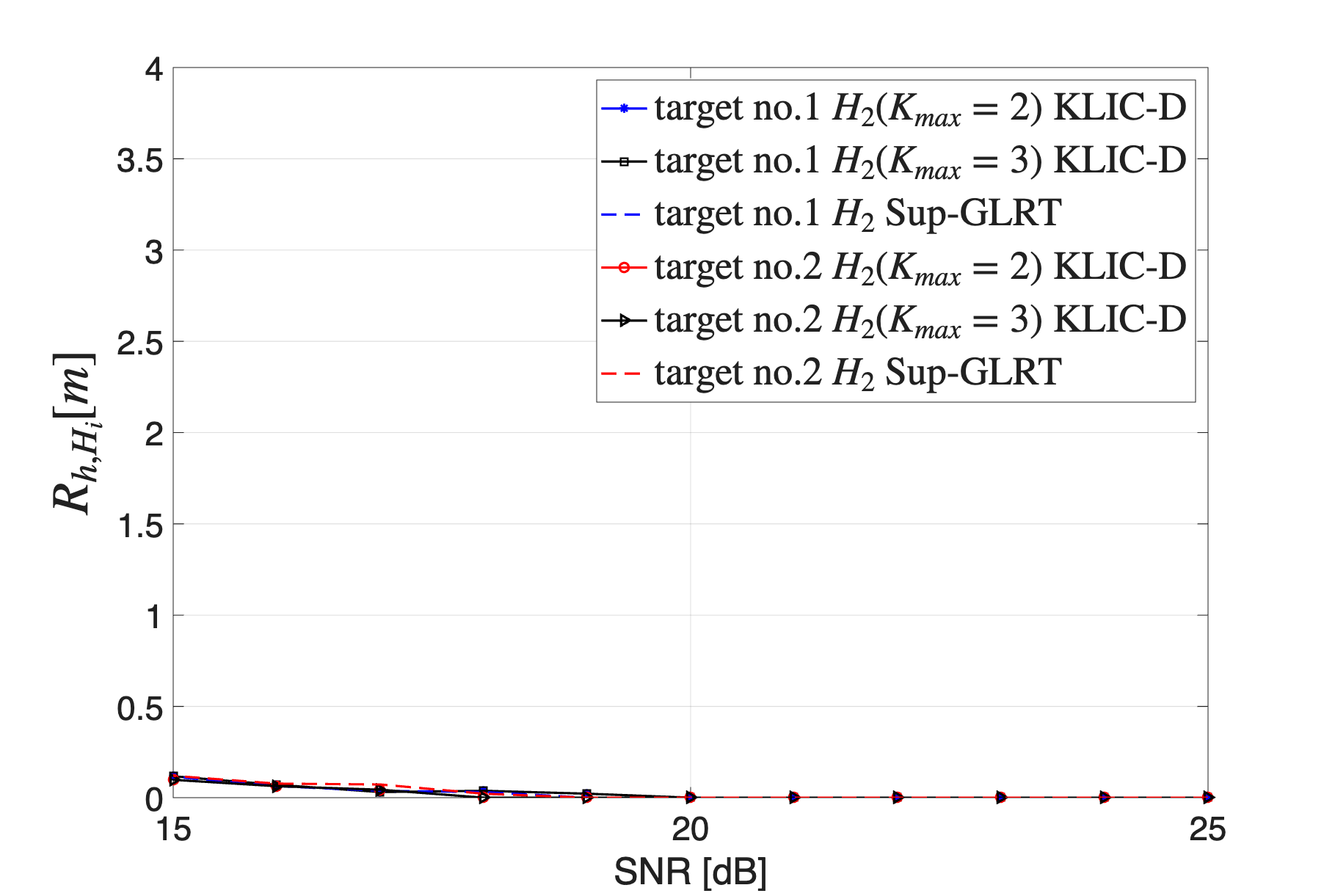}
\includegraphics[width=\linewidth]{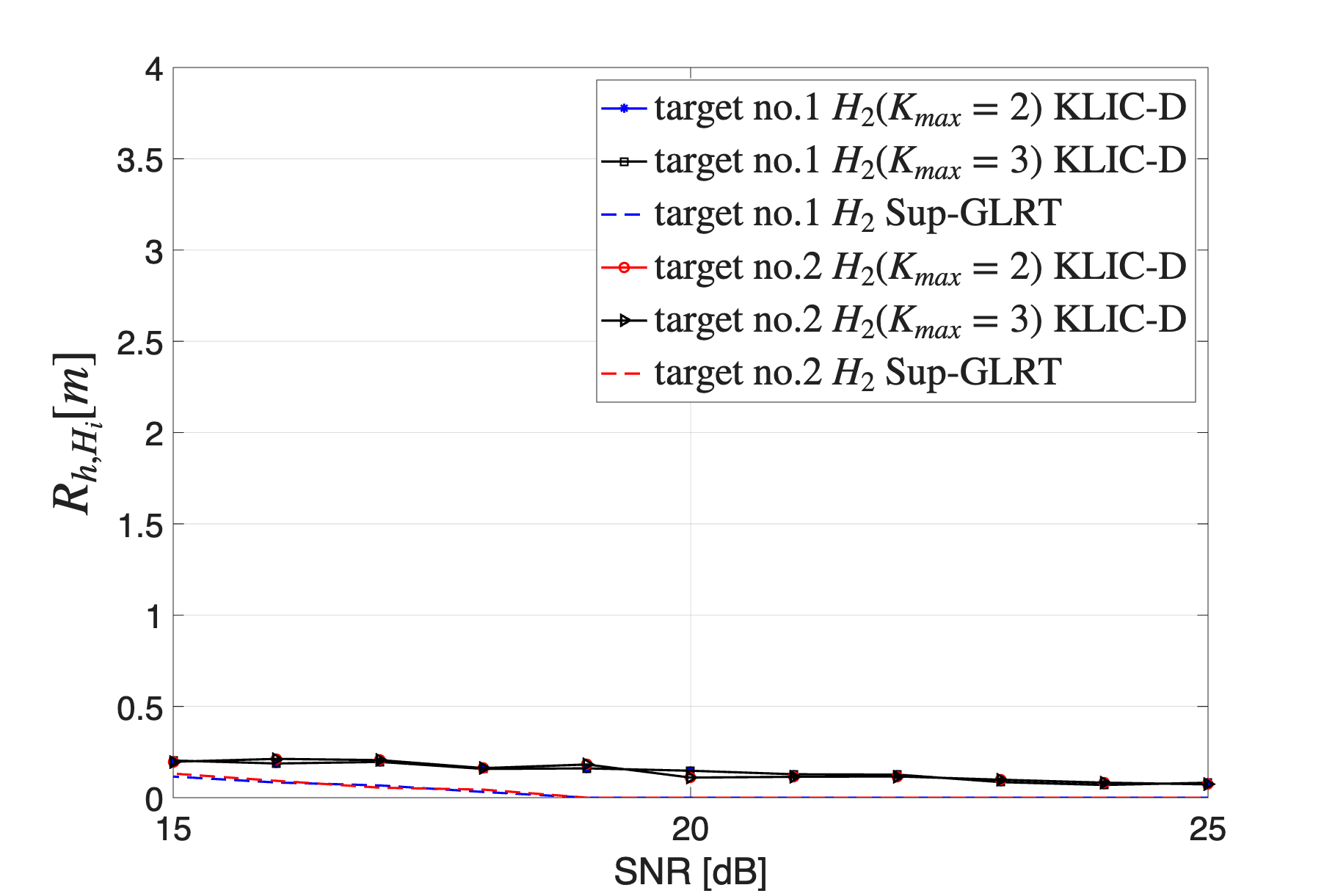}
\includegraphics[width=\linewidth]{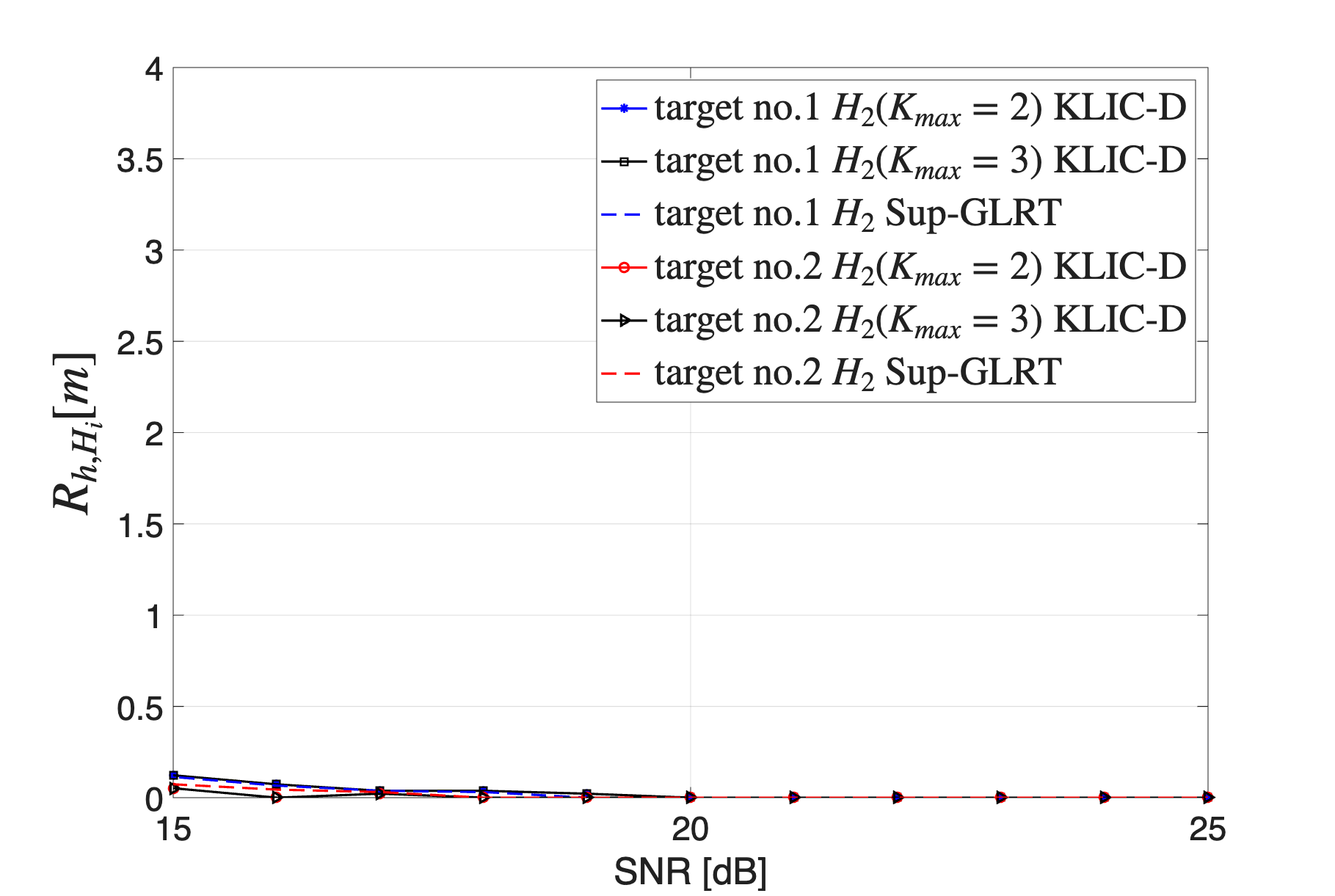} 
    \caption{RMSE for height estimation [m] versus SNR under $H_2$, assuming the same scattering coefficients with fixed positions (at the top), random positions (at the middle), different scattering coefficients with fixed positions (at the bottom).}
    \label{fig:newRMSEH2a}
\end{figure}

\begin{figure}[tbp]
    \centering
\includegraphics[width=\linewidth]{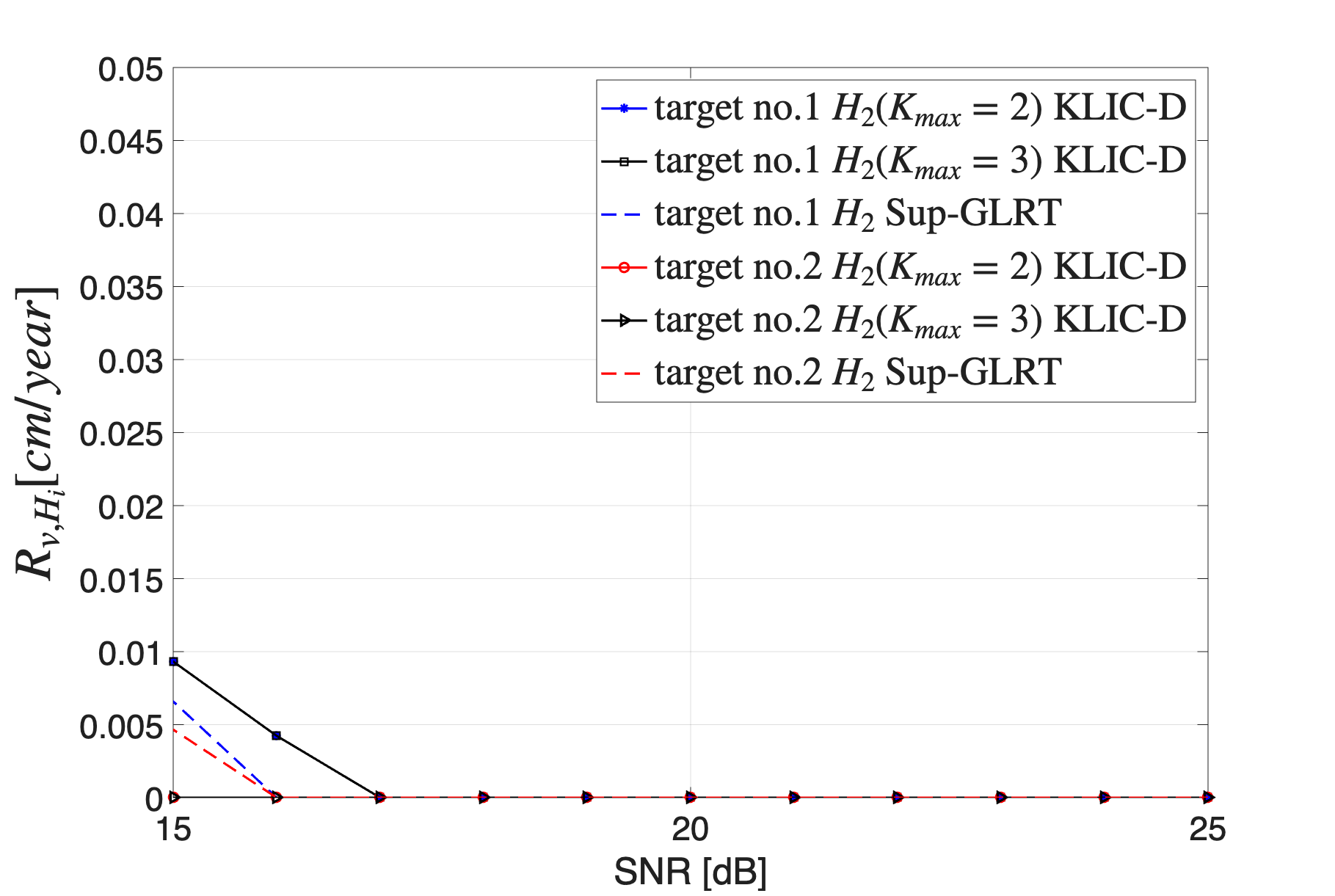}
\includegraphics[width=\linewidth]{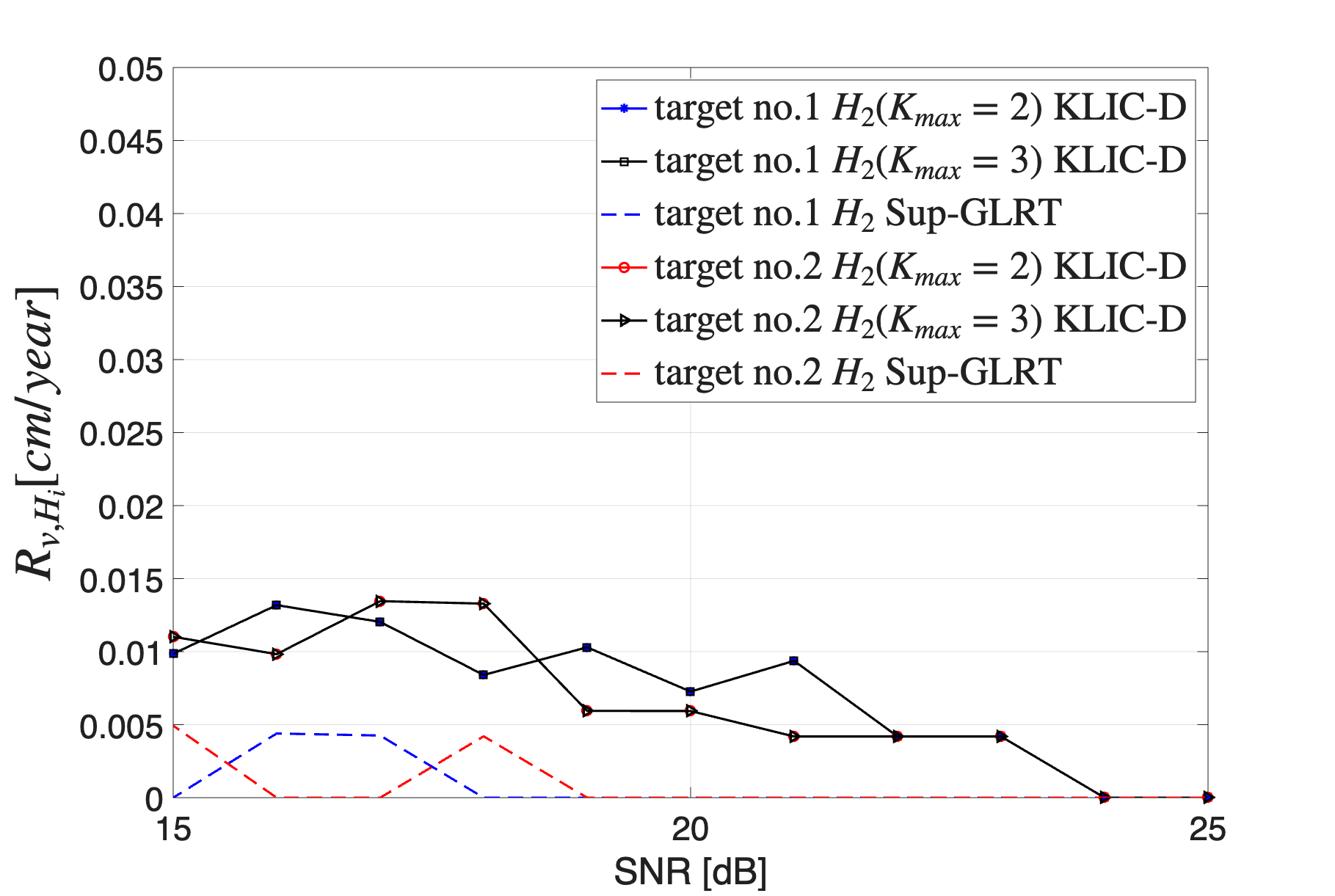}
\includegraphics[width=\linewidth]{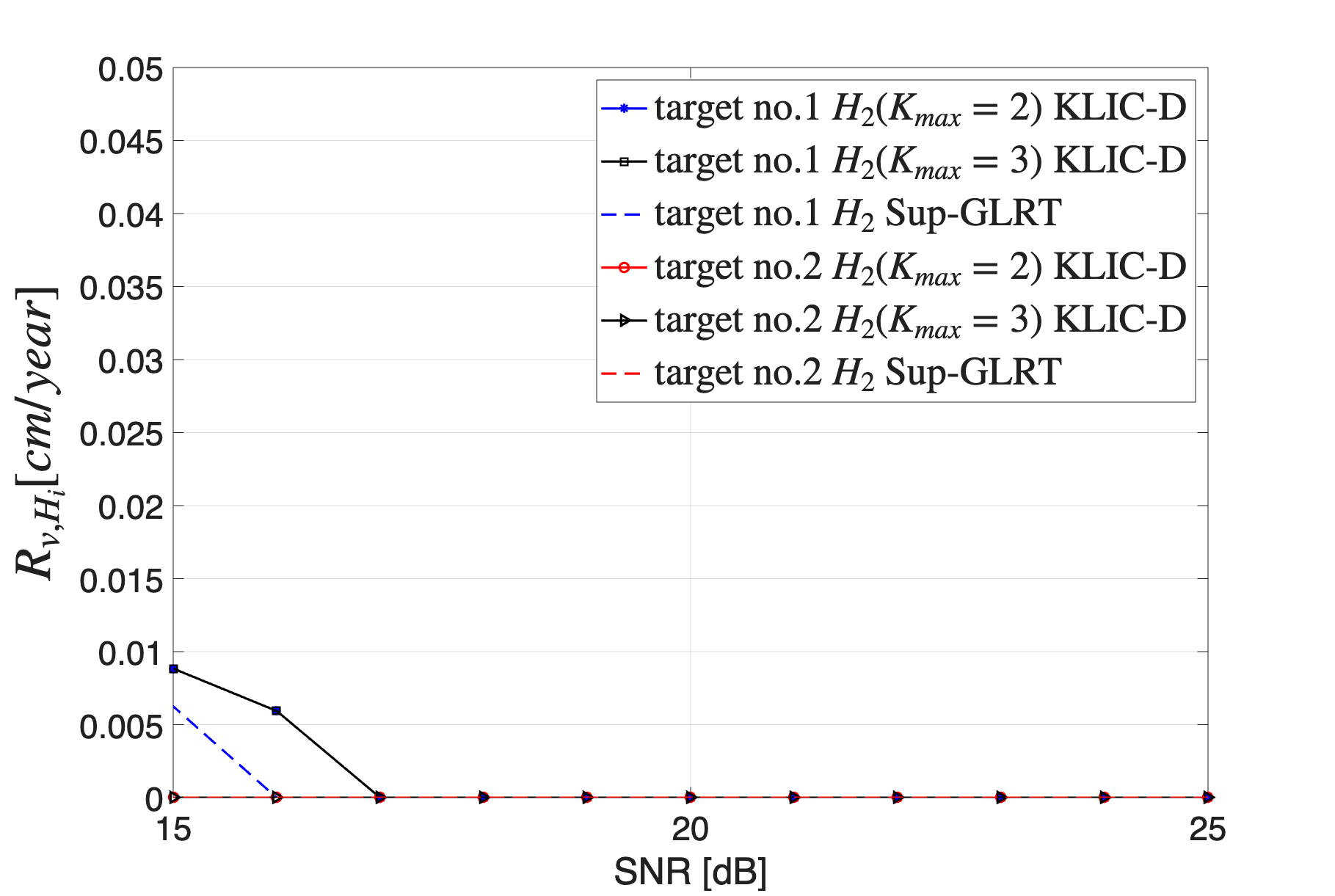} 
    \caption{RMSE for velocity estimation [cm/year] versus SNR under $H_2$, assuming the same scattering coefficients with fixed positions (at the top), random positions (at the middle), and different scattering coefficients with fixed positions (at the bottom).}
    \label{fig:newRMSEH2b}
\end{figure}

\subsubsection{Performance analysis of KLIC-D in case of $K_{\max}=3$}
In the following, the random position scenario is considered to further investigate the behavior of the KLIC-D, when $H_{i}$, $i=1,2,3$, are in force.

\begin{figure}[tbp!]
		\begin{center}
			\includegraphics[width=\columnwidth]{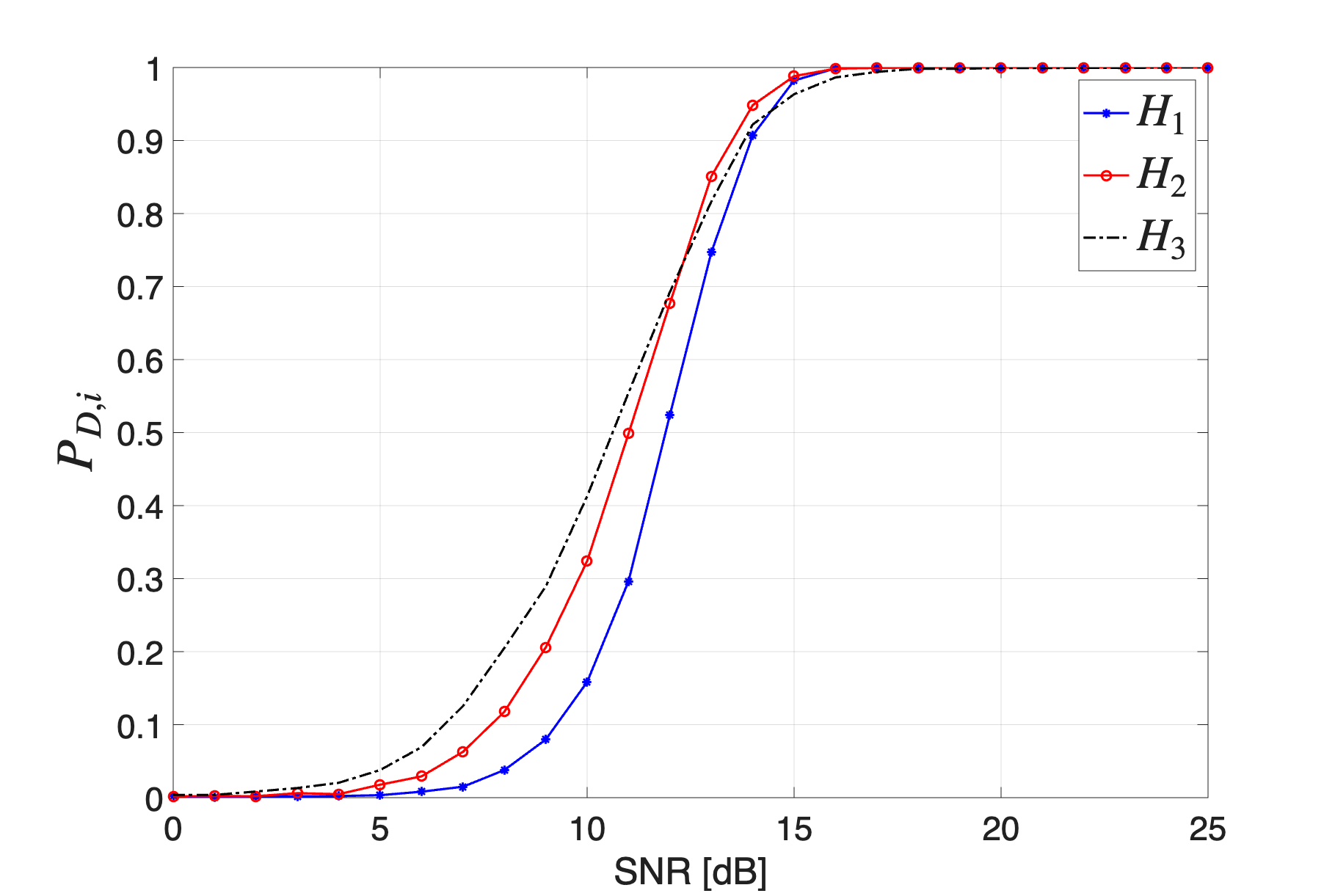}			
            \caption{{KLIC-D $P_{d,H_i}$, $i=1,2,3$ versus SNR [dB], assuming different scattering coefficients.}}
			\label{fig:P_d}
		\end{center}
\end{figure}
\begin{figure}[tbp!]
		\begin{center}
			\includegraphics[width=\columnwidth]{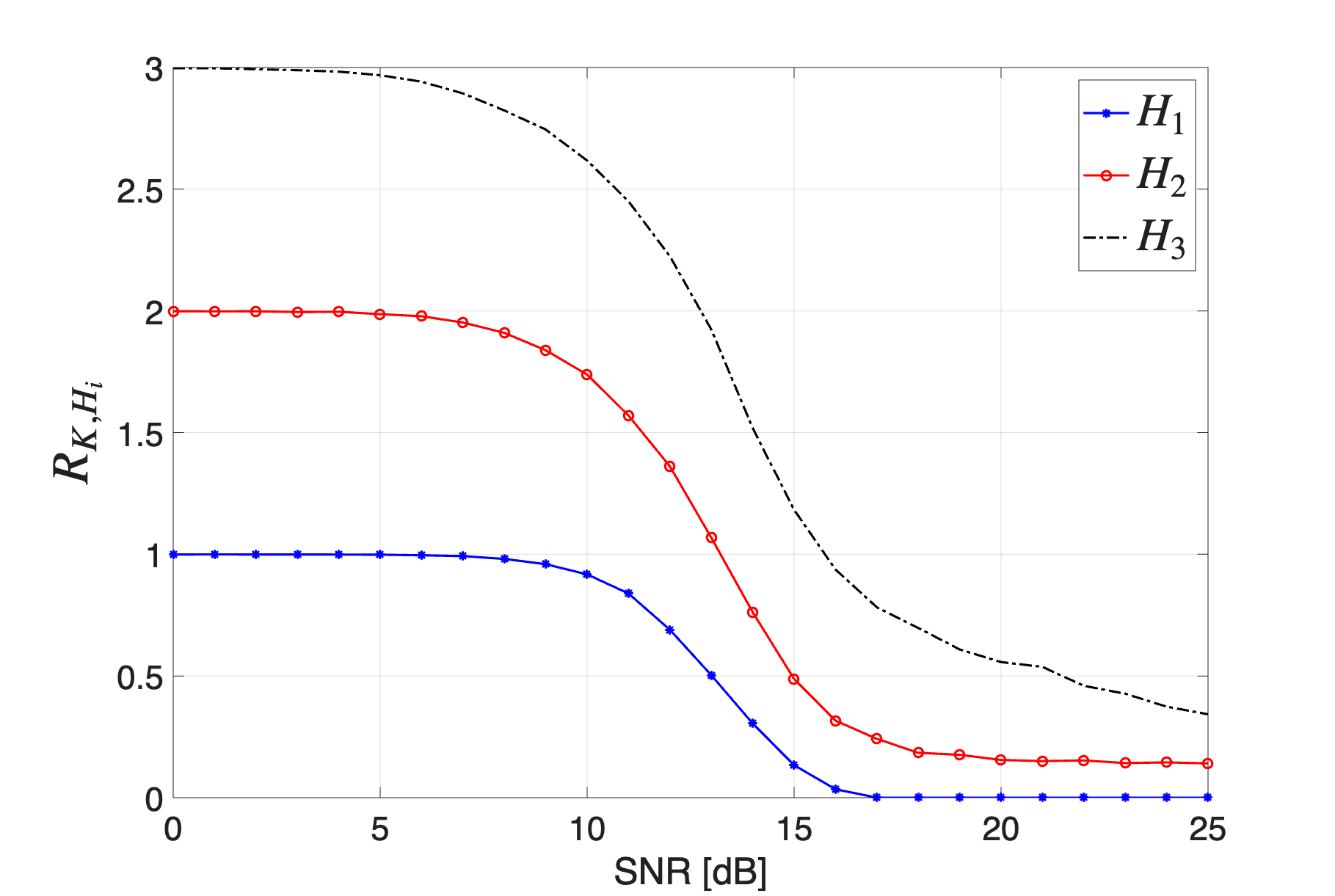}			
            \caption{{KLIC-D $R_{K,H_i}, i=1,2,3$ versus SNR [dB], assuming different scattering coefficients.}}
			\label{fig:RMS}
		\end{center}
\end{figure}


Figure \ref{fig:P_d} shows $P_{d,H_i}, i=1,2,3$ as a function of the SNR assuming different backscattering, where, the scatterers' power levels are set as follows:
\begin{itemize}
\item under $H_1$, $|g_1|^2={g^2}$;
\item under $H_2$, $|g_1|^2={g^2}$ and $|g_2|^2={1.5 g^2}$;
\item under $H_3$, $|g_1|^2={g^2}$, $|g_2|^2={1.5 g^2}$, and $|g_3|^2={2 g^2}$.
\end{itemize}
As the SNR increases, the probability of detection improves consistently. 
The detection probability curve for $H_3$ shifts to the left, indicating improved detection due to the contribution of the stronger scatterer. This behavior indicates that stronger scatterers enhance the overall detection performance, allowing the decision scheme to distinguish multiple scatterers even with medium values of SNR.

Figure \ref{fig:RMS} illustrates $R_{K,H_i}, i=1,2,3$, as a function of the SNR. 
As expected, the estimation error decreases as the SNR increases, indicating that the detector provides more accurate estimates of the number of scatterers in the high SNR regime. Moreover, as the number of scatterers grows, the estimation quality decreases due to the increase of the number of unknown parameters.

\begin{figure}[tbp]
    \centering
        \includegraphics[width=\columnwidth]{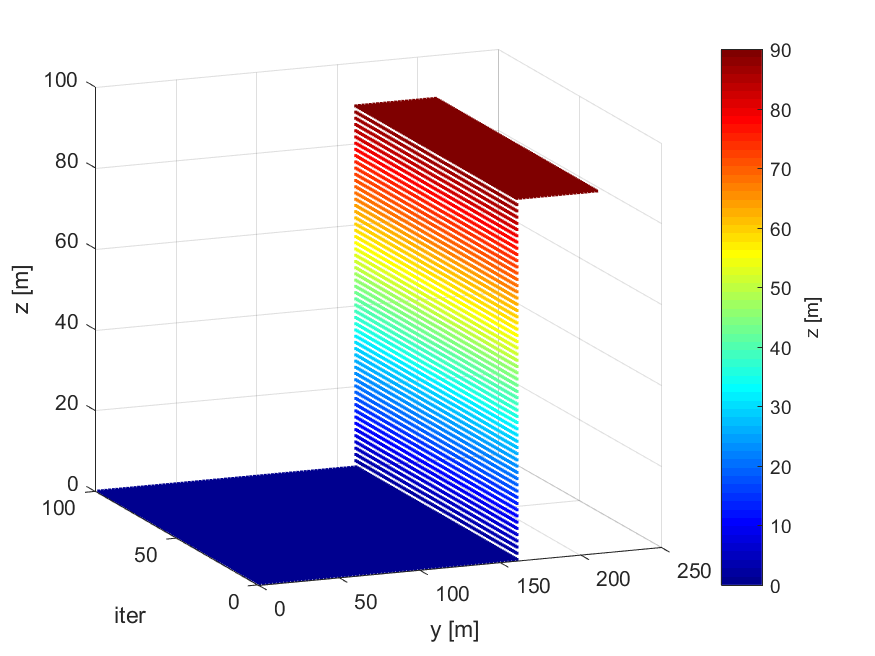}
        \includegraphics[width=\columnwidth]{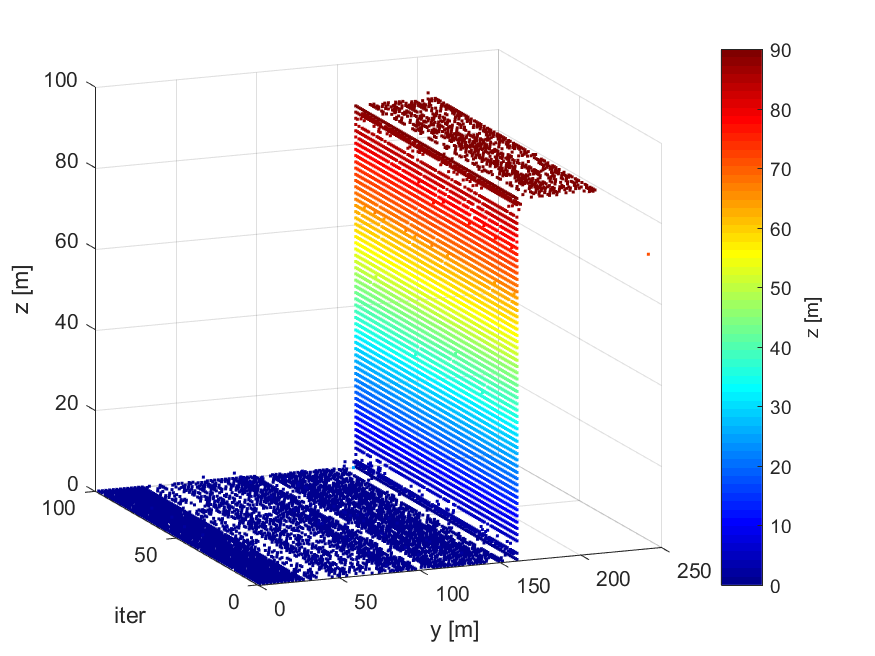}
    \caption{\textcolor{black}{3D scatter plots of the simulated scenario: (top) simulated building and (bottom) reconstruction from scatterers detected by KLIC-D with $K_{\max}=3$.}}
    \label{fig:sim_3D}
\end{figure}

{\color{black} Before concluding this section, we further consider an 
experiment in a realistic urban scenario by simulating the presence of a building with a height of approximately 89.2 m, and the surrounding floor level at $z=0$. The simulation assumes the presence of single scatterers located on the floor at near ranges, followed by pixels characterized by the interference of three scatterers from floor, façade, and roof, and, finally, in far range pixels, interference from two scatterers on the floor and façade only. We set SNR=15 dB, $g_{floor}=g_{roof}=1$, and $g_{facade}=1.5$. Received data are simulated according to the previously mentioned system parameters and baseline distributions. The values are averaged over $100$ iterations and KLIC-D uses $K_{\max}=3$. The results are shown in Figure \ref{fig:sim_3D}. It is evident that the method allows us to achieve an accurate reconstruction, including a correct handling of the triple interferences.}

A final remark on the computational load is now in order. As underlined at the end of Section III.A, the time of convergence of the compressive sensing algorithm is independent of $K_{\max}$. Moreover, experimentation on simulated data has shown that the algorithm converges in very few iterations, making the time required to estimate the scatterers parameters practically acceptable for each value of $K_{\max}$. On the other hand, Sup-GLRT estimates the scatterers parameters by scanning a multidimensional grid obtained by performing $K_{\max}$ times the Cartesian product of the search grid used for a single scatterer with itself, making it practically unfeasible for $K_{\max}\geq 3$.



\section{Experimental Results on Real data}
\label{sec_performance_real}
\begin{figure}[tbp]
    \centering
    \includegraphics[trim={1cm 1cm 2.5cm 0},clip, width=\columnwidth]{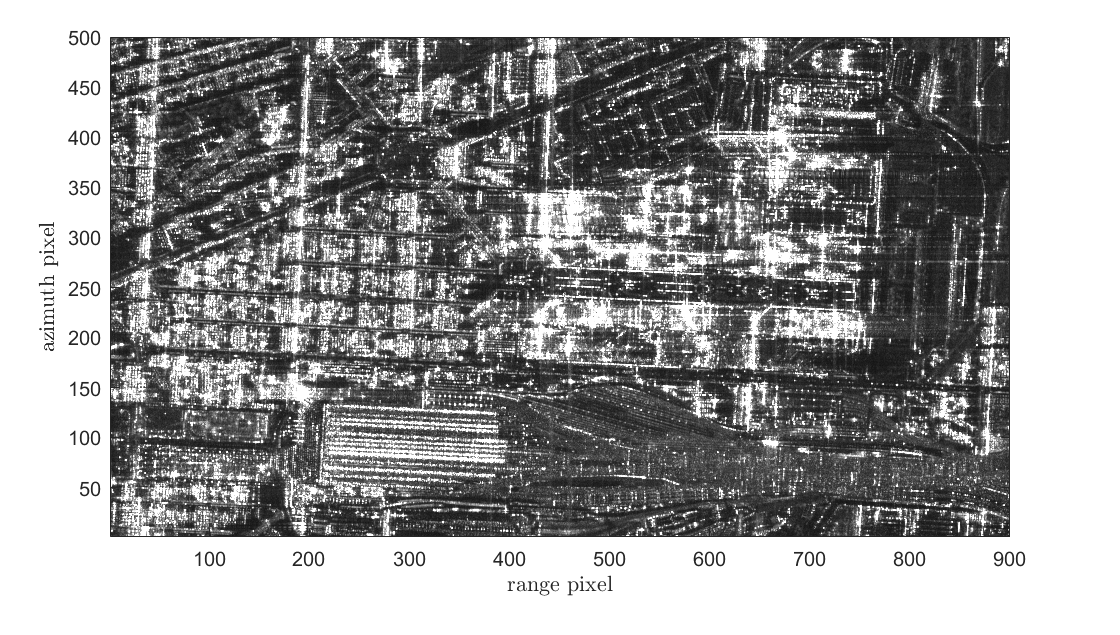}
    \caption{Temporal average amplitude image of the COSMO-SkyMed dataset over the city of Naples, South Italy. Horizontal and vertical directions corresponds to range and azimuth, respectively.}
    \label{fig:amplitude_CENTRO_DIREZIONALE}
\end{figure}

\begin{figure}
    \centering
    \includegraphics[trim={0 0 0 4.5cm}, clip, width=0.9\columnwidth]{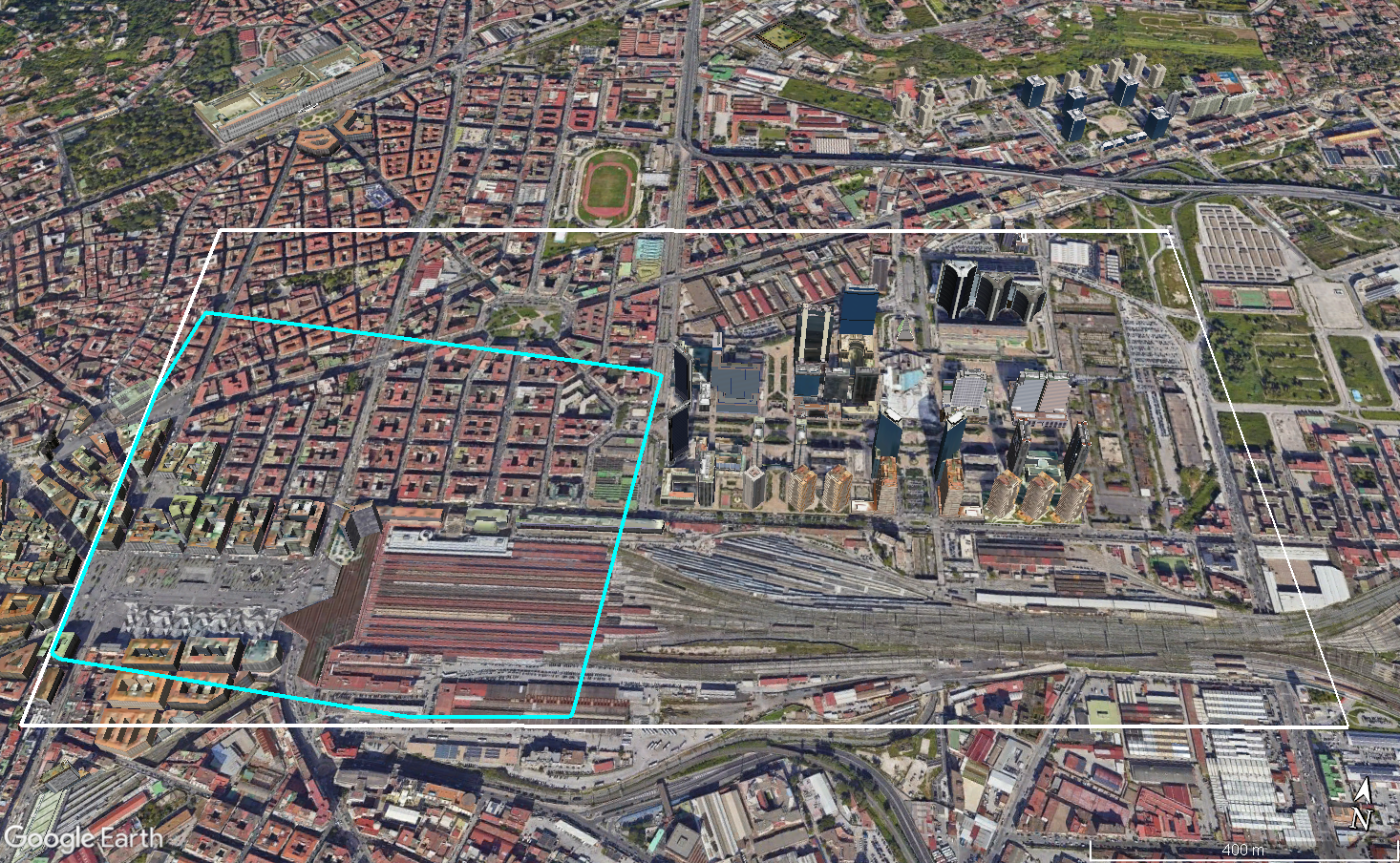}
    \caption{\textcolor{black}{Tilted view from Google Earth corresponding (white frame) to the radar image in Figure \ref{fig:amplitude_CENTRO_DIREZIONALE}. In light-blue the area corresponding to the zoom of the results in subsequent Figure \ref{fig:geo_KLIC_kmax3_zoom}.}}
    \label{fig:optical_CENTRO_DIREZIONALE}
\end{figure}

To test the effectiveness of 
the proposed detection strategy 
in real scenarios, we apply the 
proposed detector on a real 
dataset relevant to an 
urbanized environment and 
consisting of 38 SAR images 
acquired by the COSMO-SkyMed 
constellation over the city of 
Naples, South Italy. Data have 
been acquired from January 2017 
to September 2019, on ascending 
passes (HIMAGE mode) of the 
satellites. The acquisition 
system parameters as well as 
the distribution of the 
baselines are the same already 
exploited also for the 
simulations in the previous 
section, see Figure 
\ref{fig:plot_baseline_Centro_Direzionale}.
The temporal average amplitude 
of the selected test site is 
provided in Figure 
\ref{fig:amplitude_CENTRO_DIREZIONALE} \textcolor{black}{, whereas the corresponding optical image from Google Earth is shown in Figure \ref{fig:optical_CENTRO_DIREZIONALE}}: this area contains 
the city central station, 
whose platforms and tracks are 
easily recognizable in the 
bottom part of the images, as 
well as the business 
district, in the central part, 
characterized by the presence 
of a cluster of skyscrapers, 
most of them about 100 m tall, 
which lead to a frequent occurrence of layover effects.


The dataset was preliminary phase-calibrated with respect to the contributions associated with the atmospheric phase screen and large-scale deformations through the implementation of the procedure described in \cite{6832830} which first exploit a Small-Baseline processing \cite{berardino2002new} at a lower spatial resolution.
Data have been also referenced to an external Shuttle Radar Topography Mission (SRTM) Digital Elevation Model (DEM) \cite{SRTM}.

\begin{figure*}[t]
    \centering
    \includegraphics[width=0.49\textwidth]{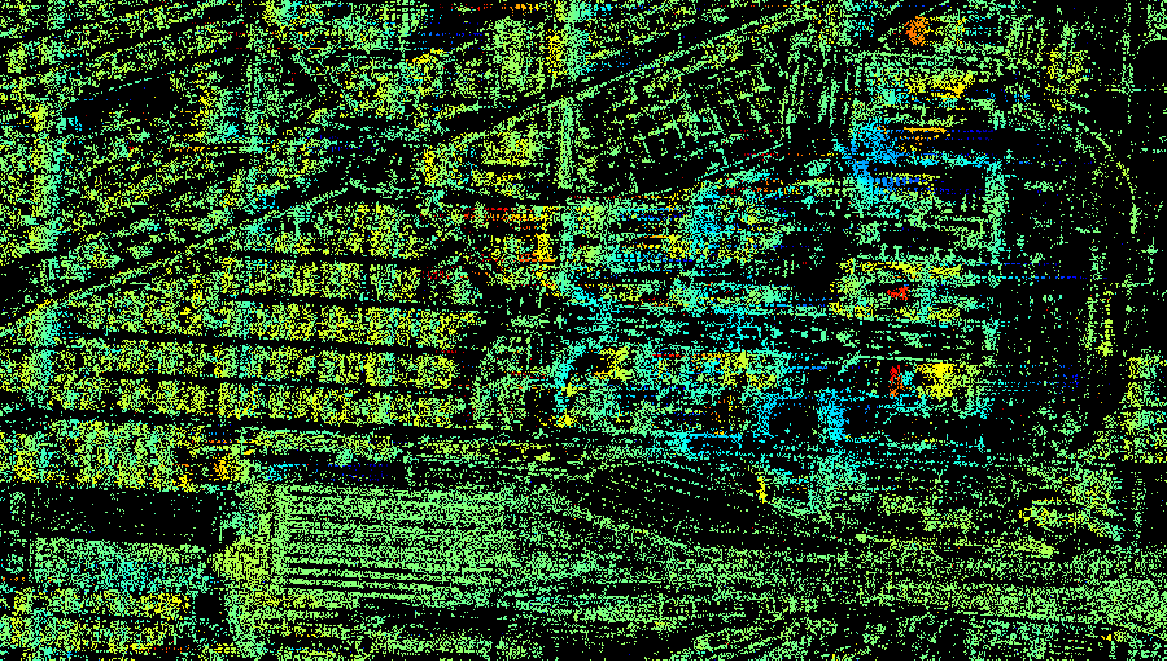}
    \includegraphics[width=0.49\textwidth]{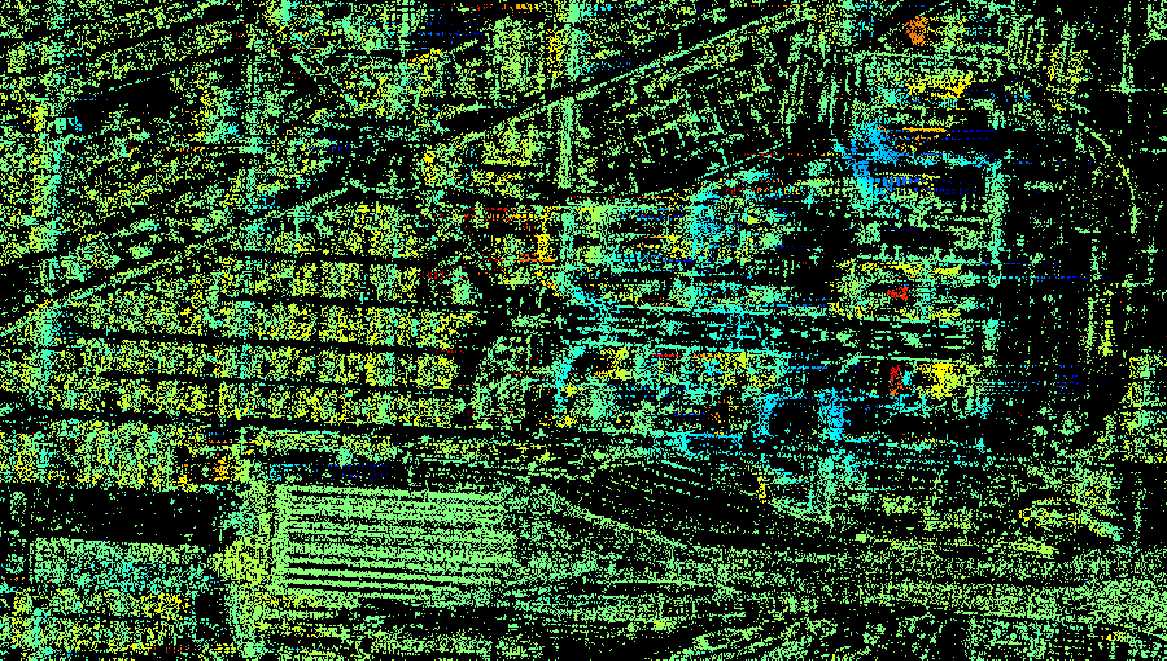}
    \includegraphics[width=0.35\textwidth]{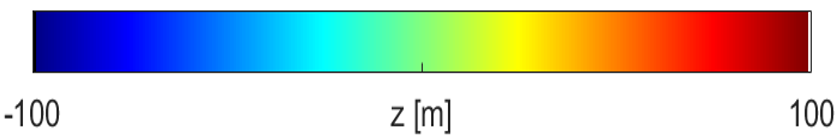}
    
    \includegraphics[width=0.49\textwidth]{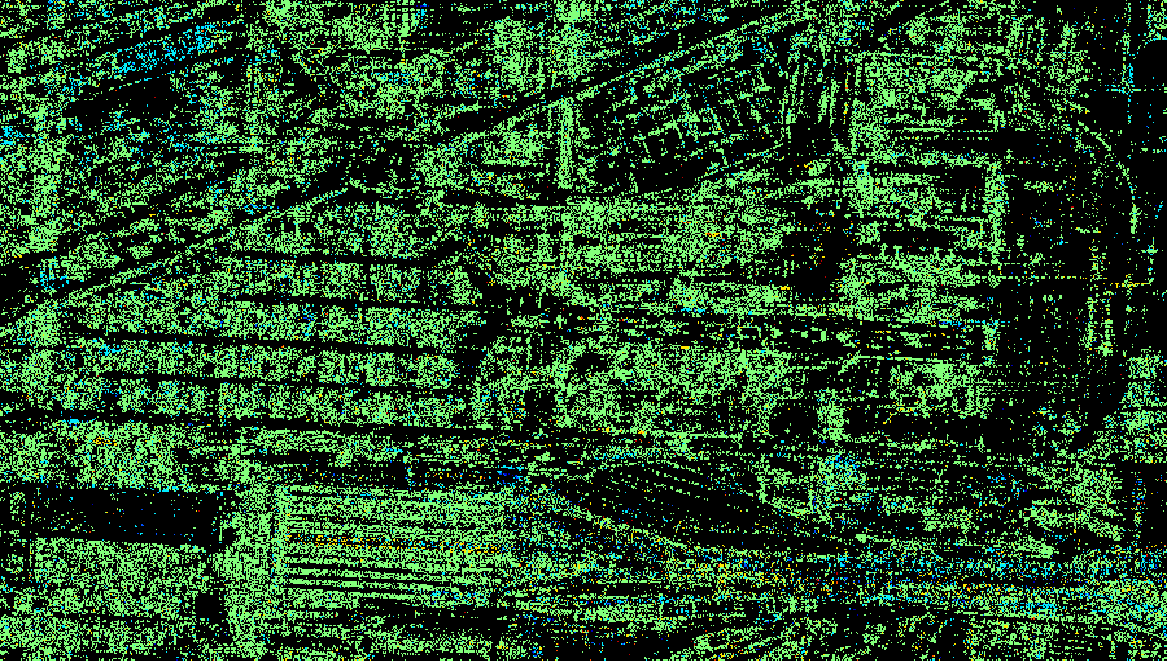}
    \includegraphics[width=0.49\textwidth]{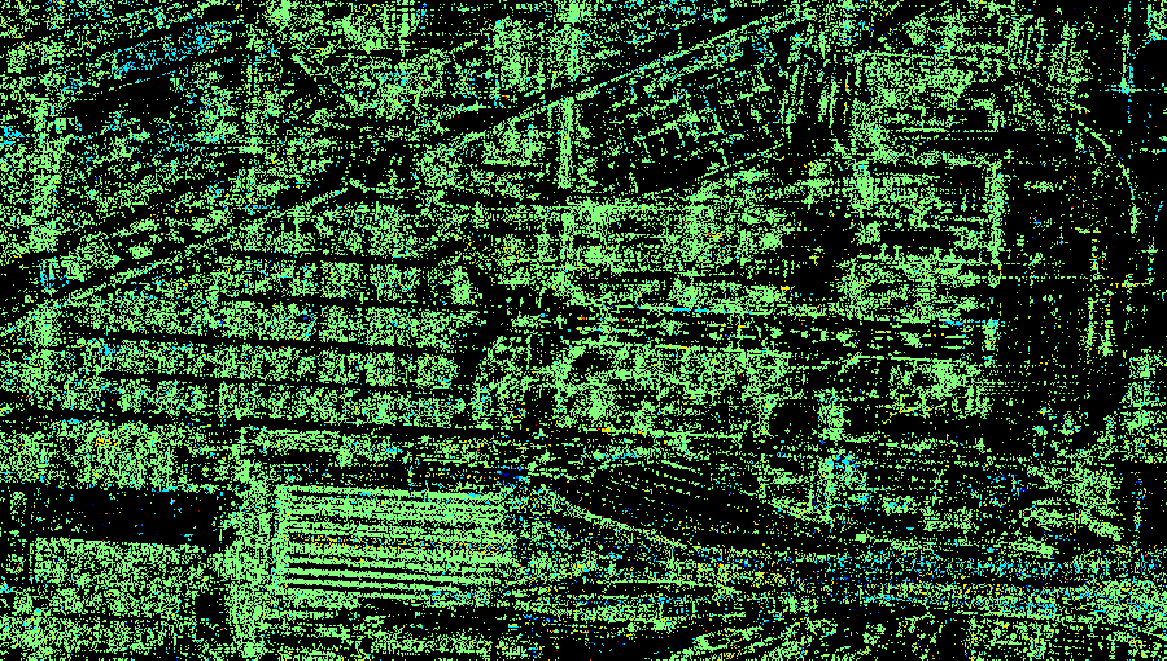}
    \includegraphics[width=0.35\textwidth]{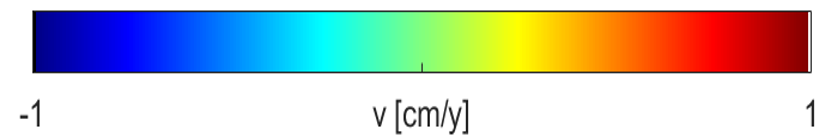}
    
    \caption{\textcolor{black}{Spatial distribution of detected single scatterers by (left column) Sup-GLRT and (right column) KLIC-D, in case of $K_{\max}=2$. (First row) estimated residual topography, (second row) estimated velocity. Colormap is set according to the values of the estimated parameters.}}
    \label{fig:res_single_scatt}
\end{figure*}

Following the same line of reasoning as in the previous section, we firstly compare the performances of the KLIC-D with Sup-GLRT by limiting the maximum number of expected scatterers to $K_{\max}=2$. The parameters of the processing, specifically, the search grids and thresholds are set according to the performance analysis in Section IV. 
The distribution of the detected single scatterers is provided in Figure \ref{fig:res_single_scatt}, where the estimated residual topography $z$, with respect to the exploited DEM, and velocity $v$, with respect to the calibrated large-scale deformation, are represented.  
The results show a very good 
agreement of the two detectors 
both in terms of number of 
detected single scatterers and 
in terms of estimated 
parameters. The estimated 
residual topography shows 
typical height ramps in 
correspondence of skyscraper 
footprints whereas (residual) 
velocities do not show any 
significant localized 
deformation. The histograms of 
the differences of the 
estimated parameters over the 
mask of overlapping detected 
pixels, normalized to the 
sampling step, set as half of 
the the theoretical Rayleigh 
resolutions for both elevation 
and velocity, show that the two 
methods estimate the same 
height over more than the \textcolor{black}{$94.4\%$} 
and the same velocity over the 
\textcolor{black}{$97.8\%$} of the common detected 
pixels. In any case, the 
residual difference is mainly 
confined to $\pm1$ sample. In 
terms of detection, the 
proposed method detects 
\textcolor{black}{$1.8\%$} more single scatterers 
than Sup-GLRT, see Table \ref{table:detected_scatterers_kmax2}. 

\begin{figure}[tbp]
    \centering
    \includegraphics[width=\columnwidth]{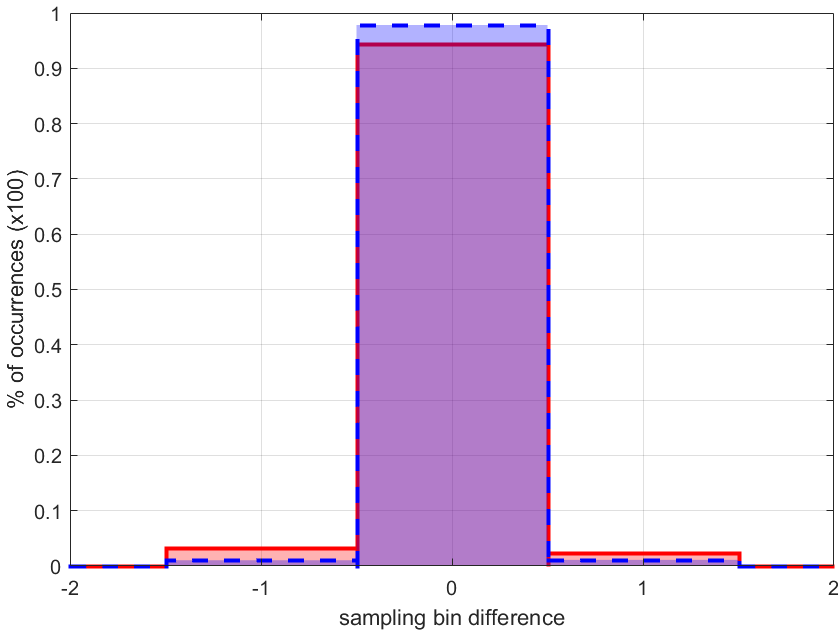}
    \caption{\textcolor{black}{Histogram of estimation differences provided by the Sup-GLRT and the KLIC-D for (red, solid line) height and (blue, dashed line) velocity. Horizontal axis is normalized to the corresponding sampling step.}}
    \label{fig:1scatt_hist_difference}
\end{figure}

\begin{figure*}[t]
    \centering
    \includegraphics[width=0.49\textwidth]{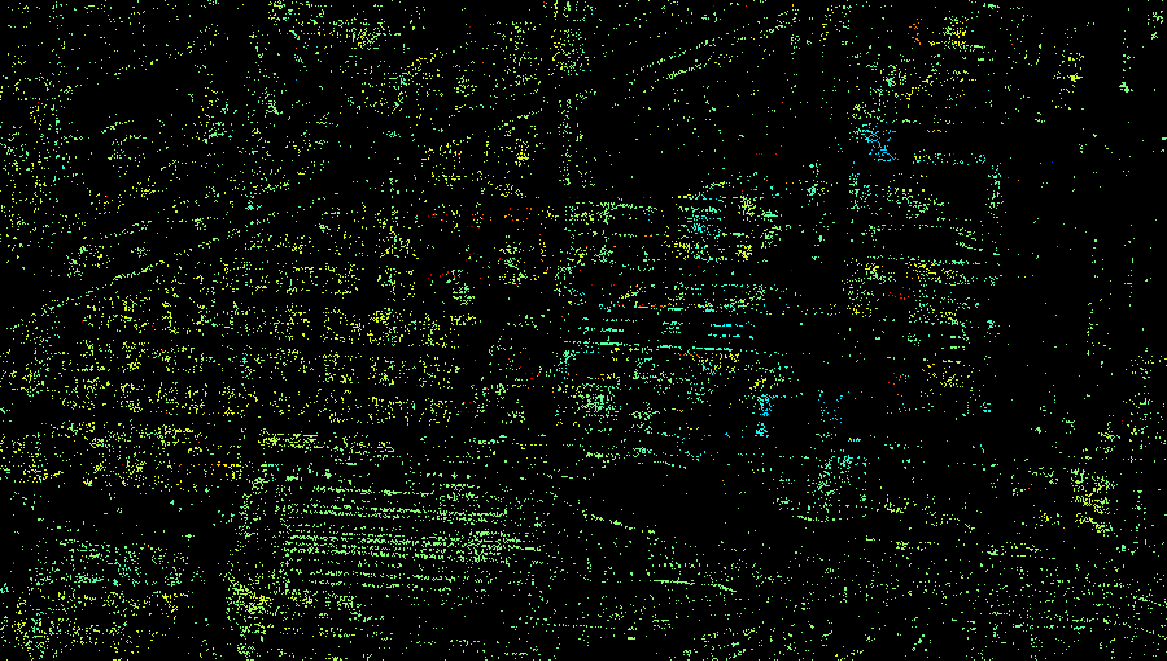}
    \includegraphics[width=0.49\textwidth]{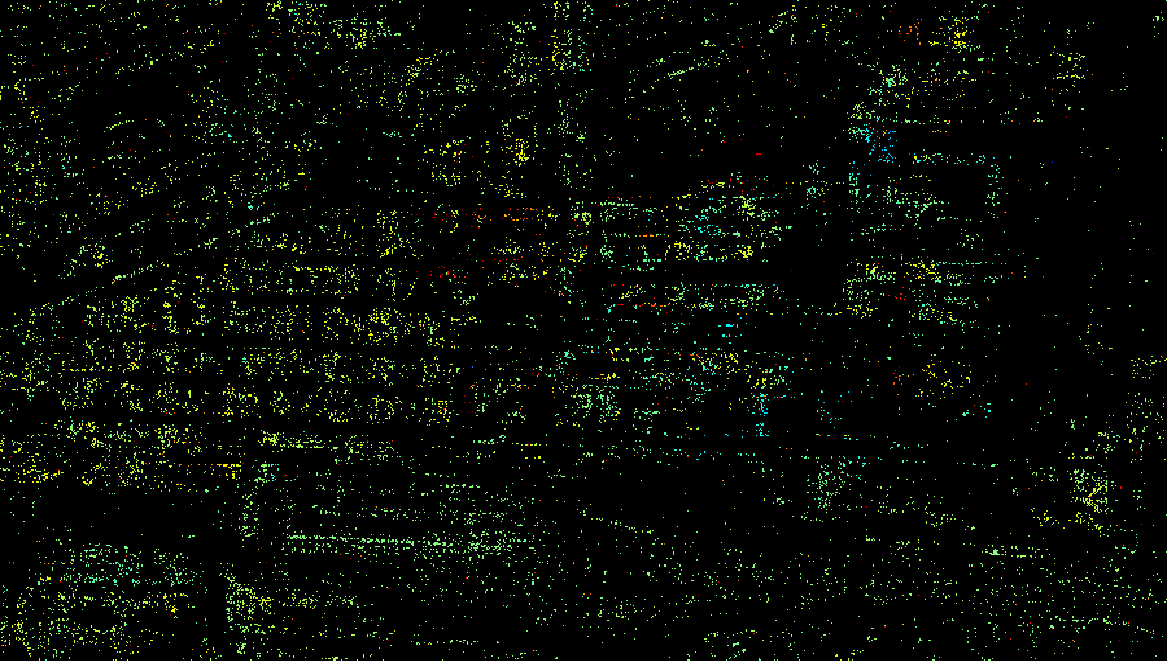}
\\ \medskip 
   \includegraphics[width=0.49\textwidth]{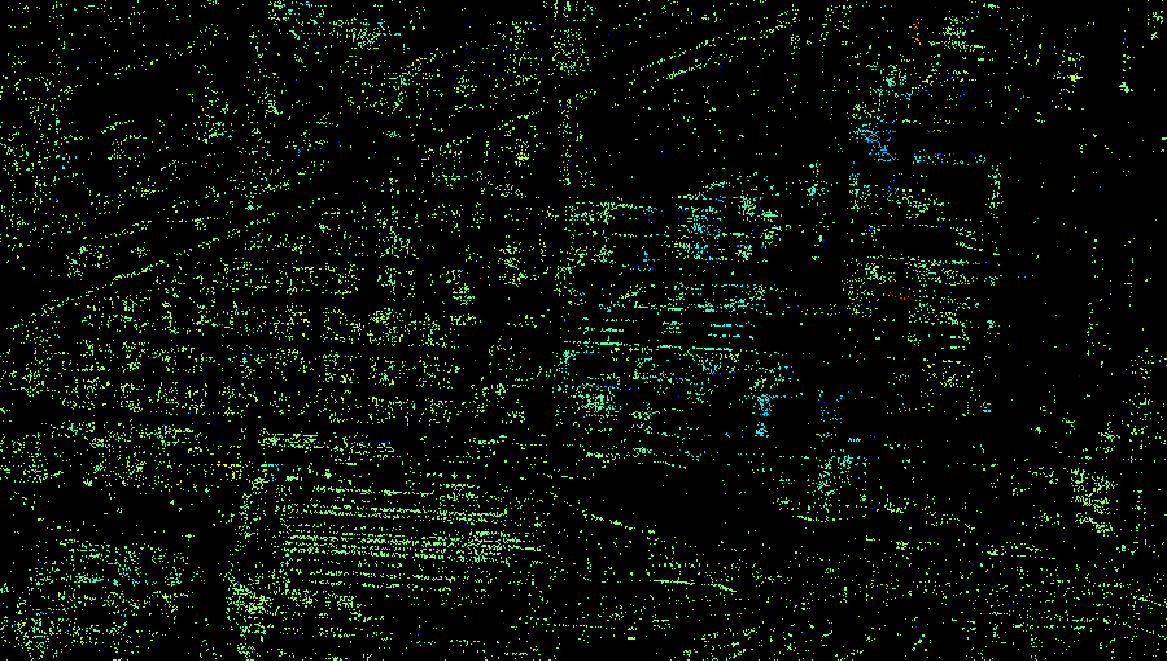}
    \includegraphics[width=0.49\textwidth]{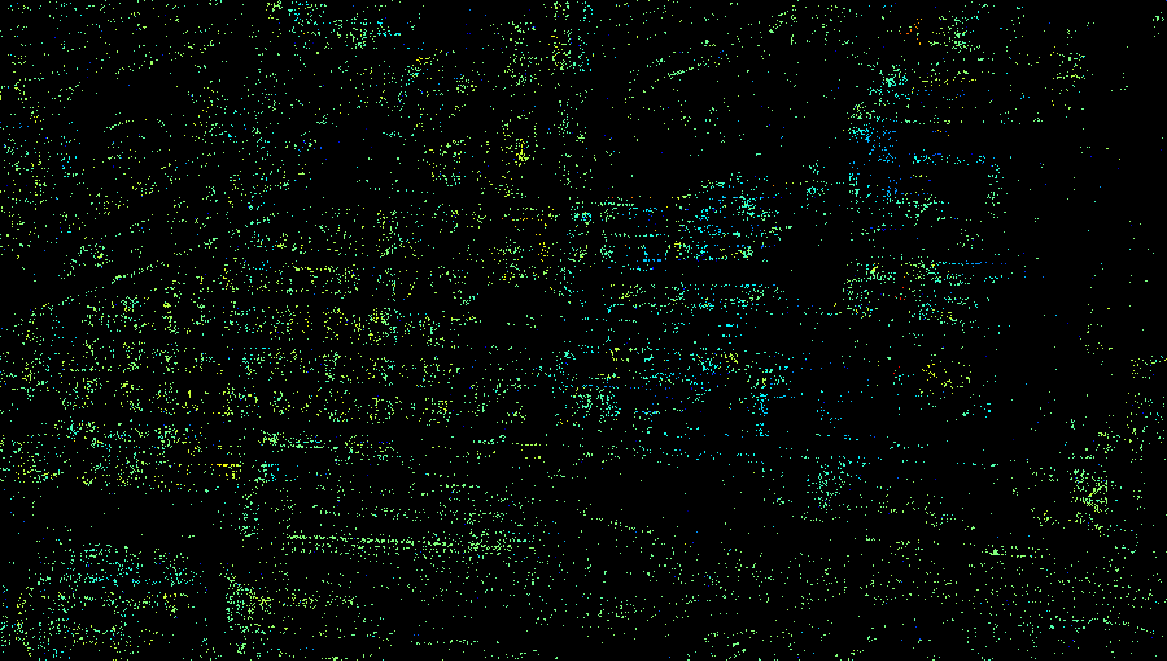}
    \includegraphics[width=0.35\textwidth]{figure_risultati_R1/colorbar_z.png}
 
    \caption{\textcolor{black}{Spatial distribution of detected double scatterers by (left column) Sup-GLRT and (right column) KLIC-D, in case of $K_{\max}=2$. (First row) higher and (second row) lower layer of estimated residual topography. Colormap is set according to the values of the estimated parameters.}}
    \label{fig:res_double_scatt}
\end{figure*}

The distribution of the 
detected double scatterers is 
provided in Figure \ref{fig:res_double_scatt}, 
where the residual topography 
associated to the higher and 
lower layers of detected 
scatterers is represented. It 
is well evident the different 
pattern for both detection 
schemes, between the higher 
level, showing more height 
variations being likely 
associated to scatterers 
located on buildings, and the 
lower level, showing a more 
uniform pattern as associated 
to scatterers lying on the 
ground. The velocity, being 
consistent with that of single 
scatterers and not showing 
significant different patterns 
between the detected double 
scatterers is not shown for 
brevity.
In terms of detection 
performances, the KLIC-D shows a reduction in the 
number of detected double 
scatterers quantified in about 
\textcolor{black}{$30\%$} than those detected by 
Sup-GLRT, see Table 
\ref{table:detected_scatterers_kmax2}. In general, by counting 
the number of pixels for which 
at least one scatterer has been 
detected, compliant with the 
definition of $P_d$ in Section 
IV, the two schemes performs 
almost the same, with a slight 
difference of \textcolor{black}{3.1\%} in 
favor of Sup-GLRT. This is 
compatible with the detection 
performances provided in Figure 
\ref{fig:P_d_comp}, where the 
two detectors shows comparable 
$P_D$s, slightly in favor of 
Sup-GLRT. Based on outcomes of 
RMSE in Figure 
\ref{fig:RMS_comp}, the 
detection loss of KLIC-D on 
double scatterers could be 
induced by the slower 
decreasing of the error with respect to SNR. 

\begin{table}[tbp]
\centering
    \caption{Number of detected scatterers - $K_{\max}=2$}
  \begin{tabular}{lccc}
   & \textbf{Sup-GLRT} & \textbf{KLIC-D} & \textbf{\% variation} \\
    \hline
    \textbf{single scatterers} & 136126 & \textcolor{black}{138562} & \textcolor{black}{+1.8\%} \\
    \hline
     \textbf{double scatterers} & 24631 & \textcolor{black}{17249} & \textcolor{black}{-30\%} \\
    \hline
     \textbf{Total} & 160757 & \textcolor{black}{155811} & \textcolor{black}{-3.1\%} \\
    \hline

  \end{tabular}
\label{table:detected_scatterers_kmax2}
\end{table}

\begin{table}[tbp]
\centering
    \caption{Number of detected scatterers - $K_{\max}=3$}
  \begin{tabular}{cc}
   $\# \text{ of PS}$ &  \textbf{KLIC-D} \\
    \hline
    \textbf{1} & \textcolor{black}{138266} \\
    \hline
     \textbf{2} & \textcolor{black}{16722} \\
    \hline
     \textbf{3} & \textcolor{black}{959} \\
     \hline
    \textbf{Total} & \textcolor{black}{155947} \\
    \hline

  \end{tabular}
\label{table:detected_scatterers_kmax3}
\end{table}

\begin{figure*}[t]
  \centering
  \begin{minipage}[t]{0.49\textwidth}
    \centering
    \includegraphics[width=\linewidth]{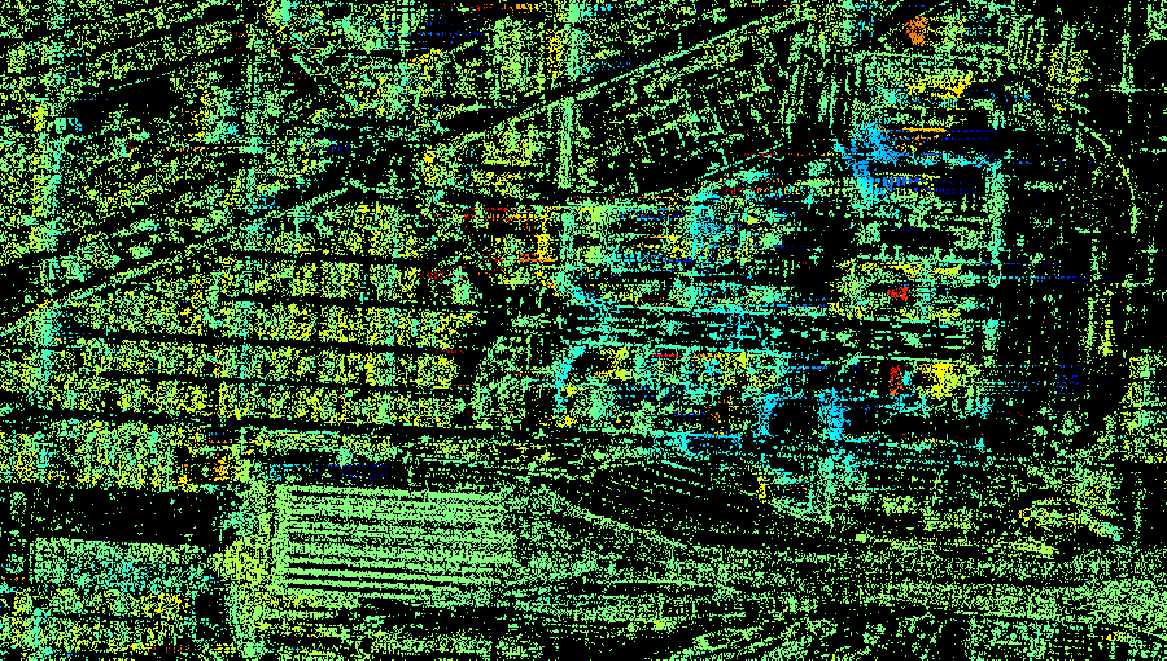}\\[1ex]
    \includegraphics[width=\linewidth]{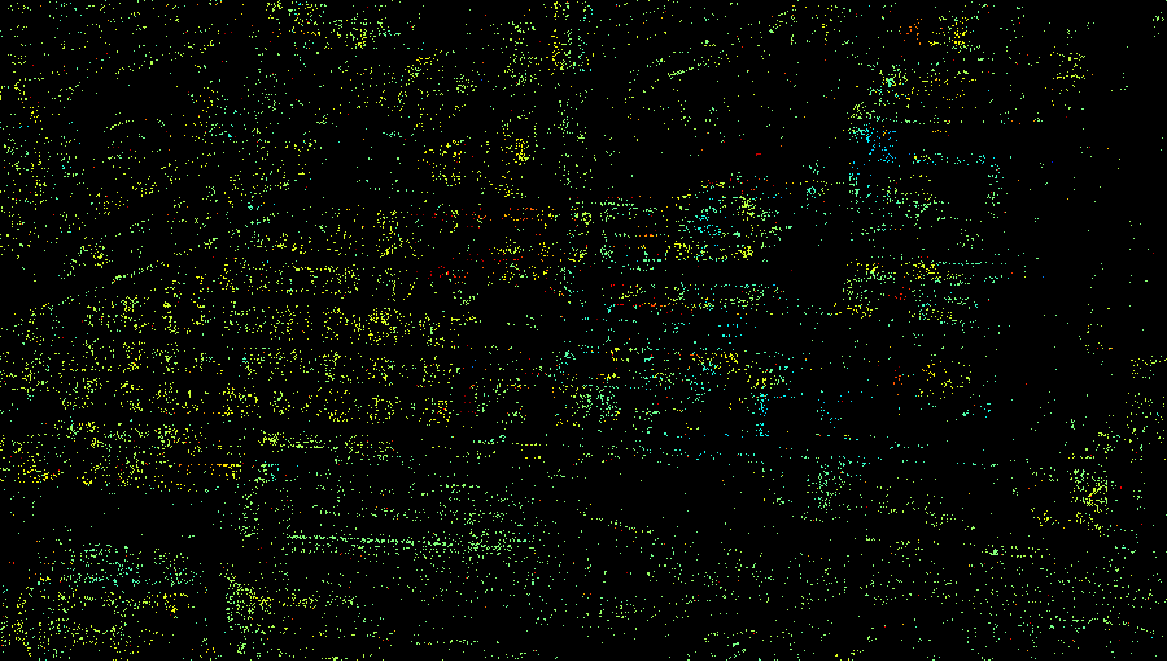}\\[1ex]
    \includegraphics[width=\linewidth]{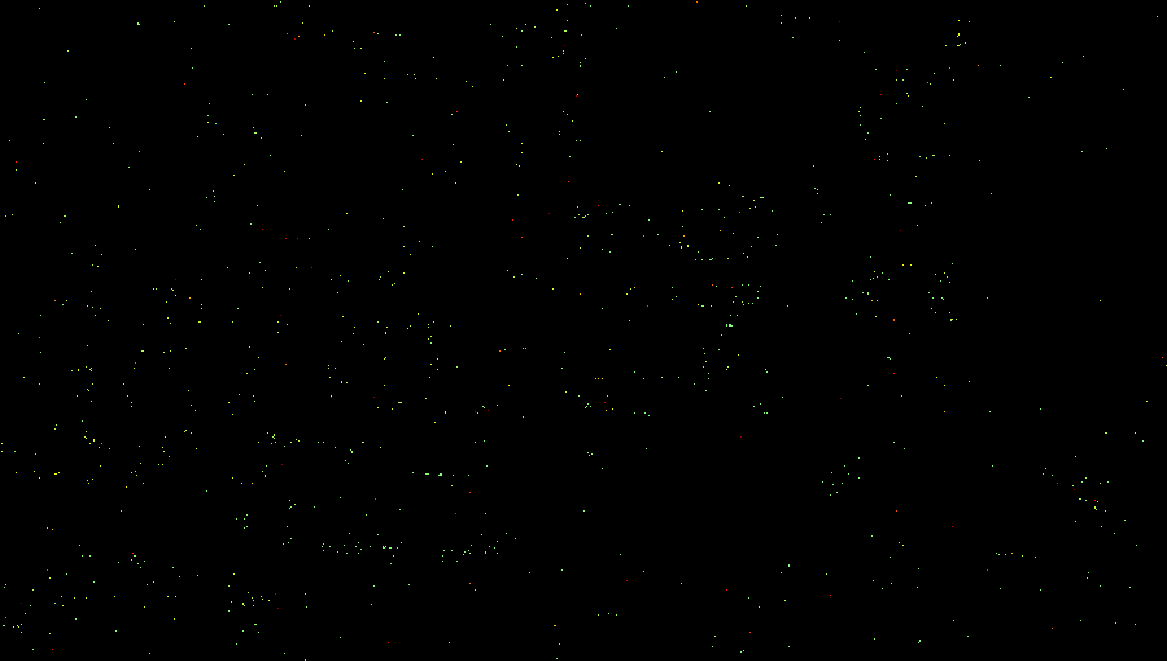}
    \includegraphics[width=0.7\textwidth]{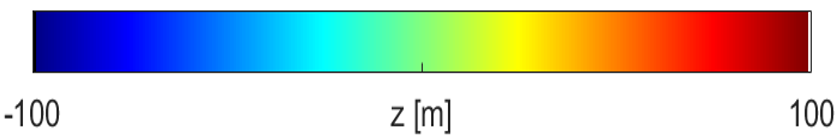}
  \end{minipage}
  \hfill
  \begin{minipage}[t]{0.49\textwidth}
    \centering
    \includegraphics[width=\linewidth]{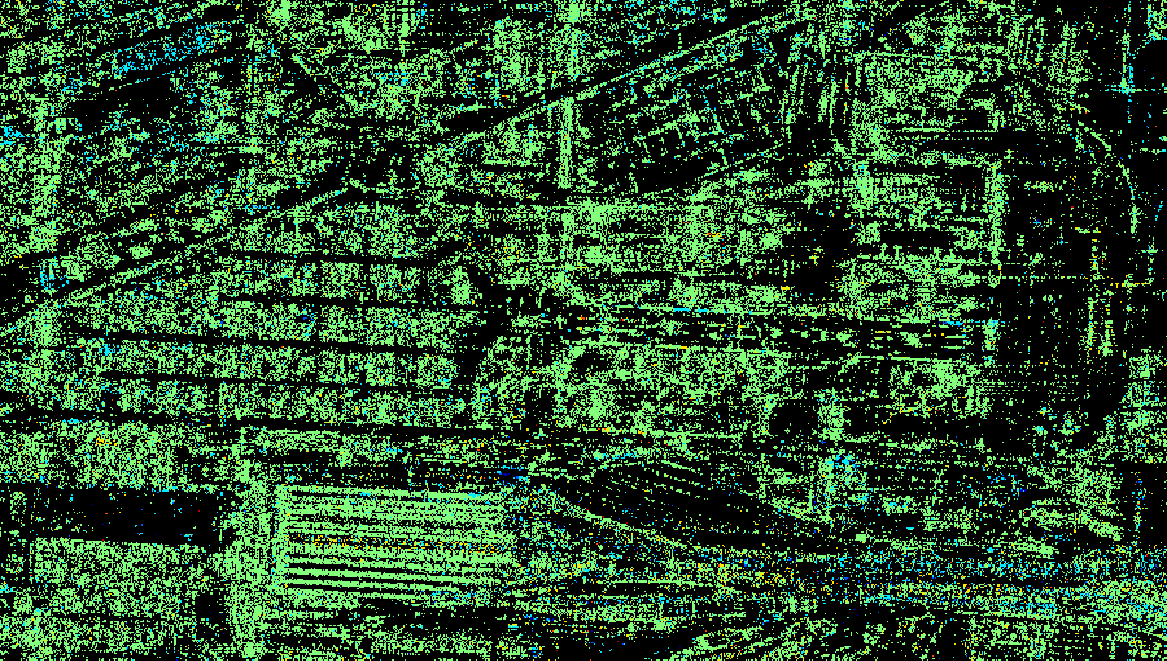}\\[1ex]
    \includegraphics[width=\linewidth]{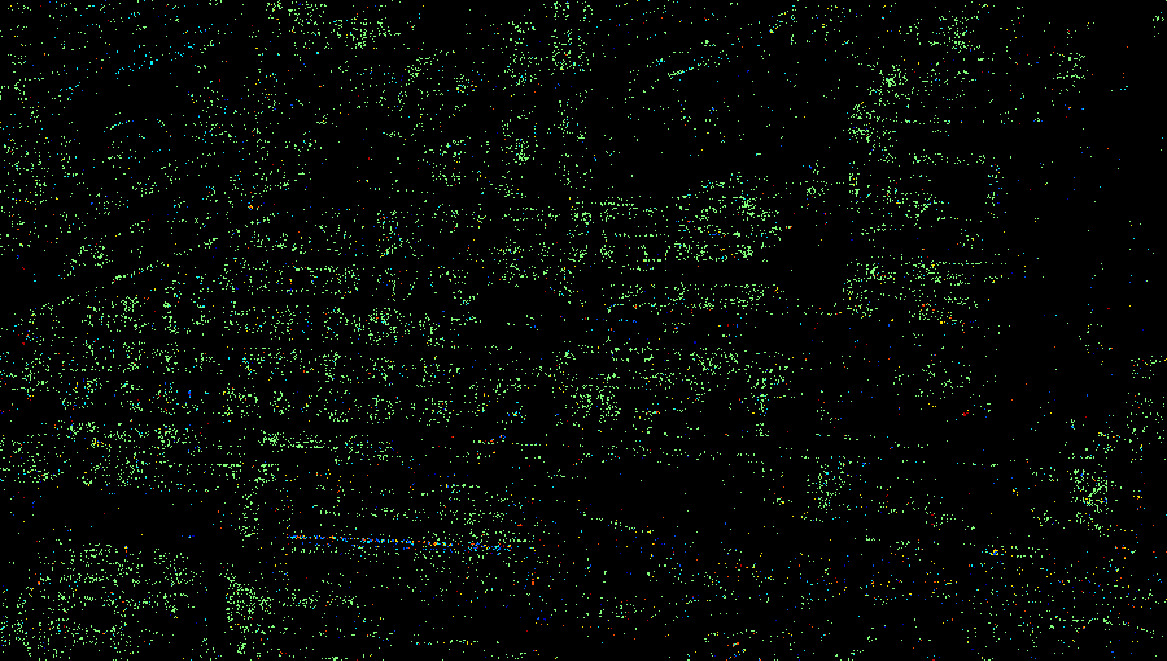}\\[1ex]
    \includegraphics[width=\linewidth]{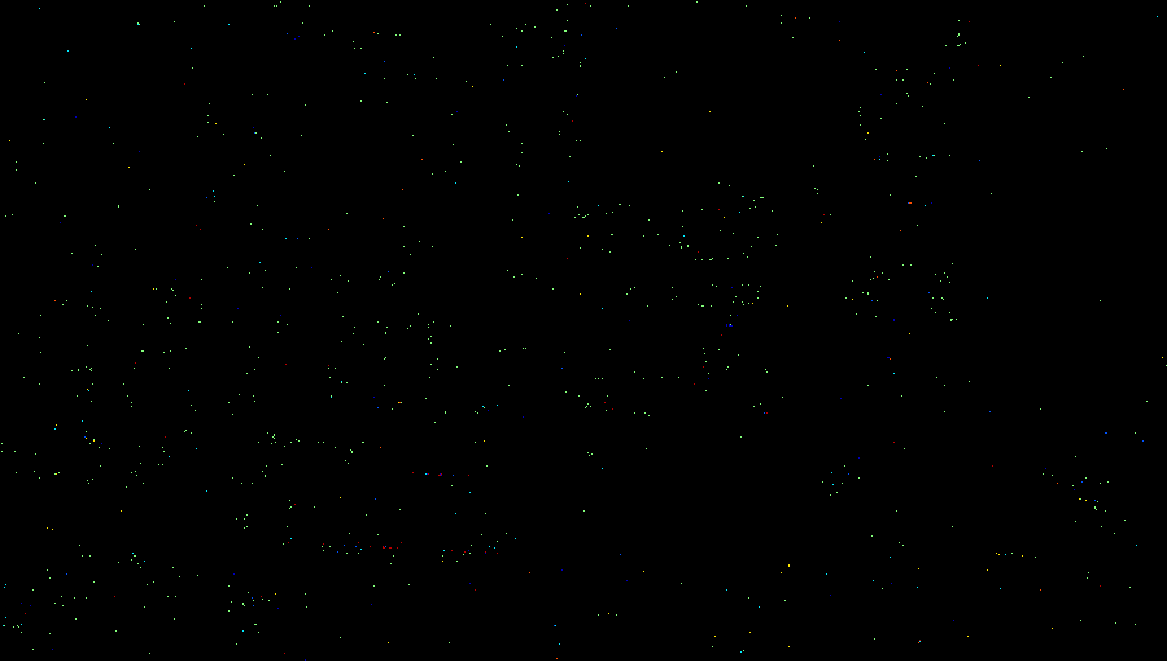}
    \includegraphics[width=0.7\textwidth]{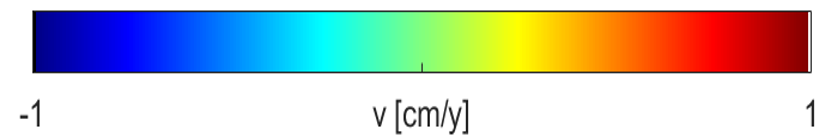}
  \end{minipage}
  \caption{\textcolor{black}{Spatial distribution of triple scatterers detected by KLIC-D for $K_{\max}=3$. Estimated (left column) residual topography  and (right) velocity for (first row) $\#\text{ of PS}=1$, (second row) $\#\text{ of PS}=2$ and (third row) $\#\text{ of PS}=3$. Colormap is set according to the values of the estimated parameters.}}
  \label{fig:detected_scatterers_kmax3}
\end{figure*}

The strength of the proposed 
method stands, however, in the 
possibility to increase the 
order of the maximum possible 
detectable scatterers without 
increasing the computational 
cost, differently form Sup-GLRT 
which cannot be extended so 
effortlessly. The results of 
the implementation of the 
proposed detector for $K_{\max}=3$ 
are provided in Figure 
\ref{fig:detected_scatterers_kmax3}, 
where the estimated residual 
topography and velocity for 
single and higher layer of 
double and triple scatterers 
are represented. The results of 
both single and double 
scatterers are in accordance 
with those of Figures 
\ref{fig:res_single_scatt} and 
\ref{fig:res_double_scatt}. 
Additionally, the estimation of 
the sparse triple scatterers do 
not show significant anomalies 
compared to the former two. In 
terms of number of detected 
scatterers, it results from 
Table 
\ref{table:detected_scatterers_kmax3} 
that the number of total 
scatterers, for which at least 
one scatterer is detected, is almost the same as for the 
case of $K_{\max}=2$. This is 
reasonable with the working 
principle of the proposed 
detector that, by increasing 
$K_{\max}$, extends the test to 
the increased number of 
potential peaks provided by the 
CS estimation.   

\begin{figure}[tbp]
    \centering
    \includegraphics[width=\columnwidth]{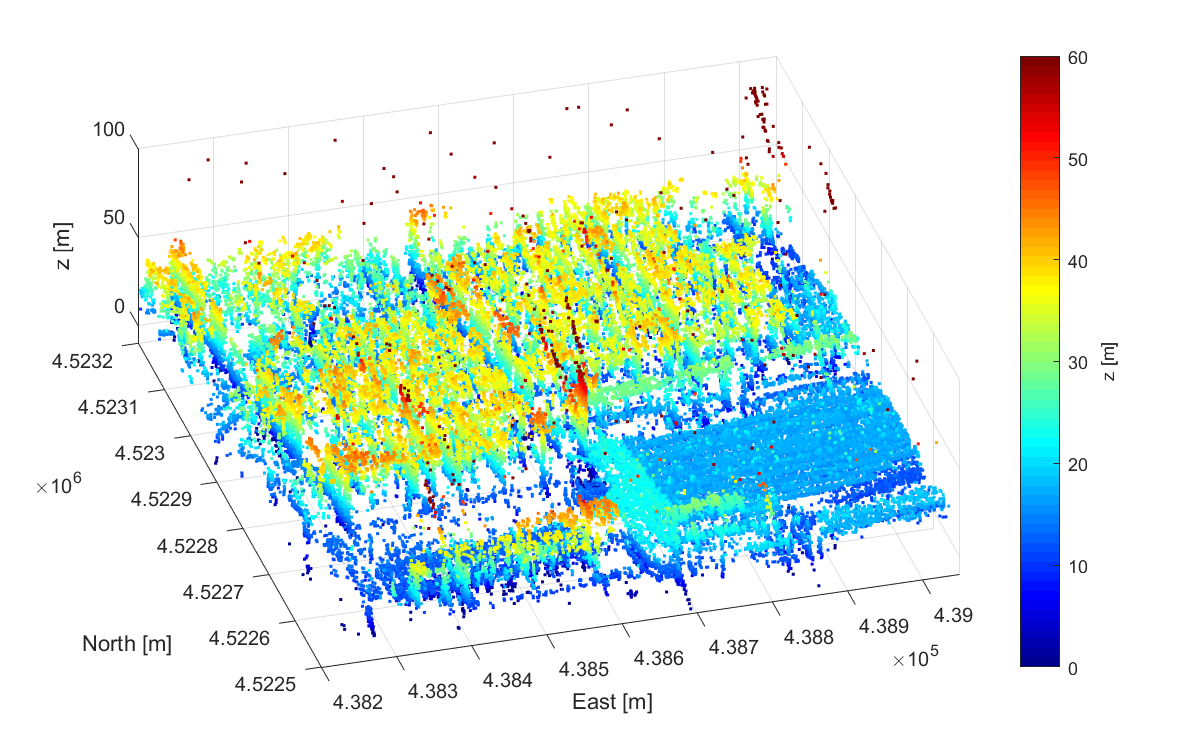}
    \caption{\textcolor{black}{3-D point cloud of scatterers detected by KLIC-D ($K_{\max}=3$) on the area delimited by the light-blue frame in Figure \ref{fig:optical_CENTRO_DIREZIONALE}. }}
    \label{fig:geo_KLIC_kmax3_zoom}
\end{figure}

\textcolor{black}{The 3-D point cloud of scatterers detected by KLIC-D with $K_{\max}=3$ in the area delimited by the light-blue frame in Figure \ref{fig:optical_CENTRO_DIREZIONALE}, after geocoding from the native radar (azimuth; range) geometry to a geographical (East; North; Height) reference system is provided in Figure \ref{fig:geo_KLIC_kmax3_zoom}. The shapes of the dense building texture can be easily recognized as well as the more uniform pattern corresponding to the station and platforms in the bottom-right part of the image. False alarms are detected in correspondence of very strong backscattering sources leading to significantly large focusing sidelobes, see also Figure \ref{fig:amplitude_CENTRO_DIREZIONALE}, although Hamming filtering was applied in the focusing process. Mitigation of those sidelobes \cite{SideLobe_Cancelation}, that would be beneficial, is however out of the scope of this work.}

\section{Conclusion}
\label{sec_conclusions}
In this work we have introduced a novel decision scheme in the framework of SAR Tomography to detect interfering scatterers within the same same resolution cell. The detector relies on an information-theoretic design framework and allows us to solve multiple hypothesis test formed by one null hypothesis and multiple alternative hypotheses. For the specific case, each alternative hypothesis correspond to a given number of scatterers. 
Remarkably, this approach leads to the same decision scheme as that proposed in \cite{DeMaio2009} when only one PS is present. 
In order to limit the computational burden, we have coupled the KLIC-based design framework proposed in \cite{addabbo2021adaptive} with the compressive sensing. Doing so, the resulting new detector exhibits a significant practical value since it does not require a number of classification/detection threshold that grows with the number of PSs unlike existing multistage approaches. At the same time, the performance assessment, carried out on both synthetic and real data, has shown that the detection/classification results are similar to that of the Sup-GLRT \cite{budillon2015glrt} that is the most natural competitor due to its scalability according to the number of PSs.

Future research tracks might encompass the assessment of a detector obtained by further iterating the estimation of the unknown parameters as well as the design under more involved interference models.
{\color{black}
Another interesting research track concerns the design of two stage architectures
formed by cascading the proposed estimation algorithm (or other grid-based efficient
solutions) and a gridless estimation method that exploits the rough estimates
provided by the first stage as initial points.
}
	
\section*{Acknowledgment}
The Authors would like to thank the Italian Space Agency (ASI) for providing the COSMO-SkyMed dataset.
	

\bibliographystyle{IEEEtran}

\end{document}